\documentclass[runningheads]{llncs}

\usepackage{type1cm}        
\usepackage{makeidx}         
\usepackage{graphicx}        
\usepackage{multicol}        
\usepackage[bottom]{footmisc}

\usepackage{newtxtext}       %
\usepackage{newtxmath}       

\usepackage{graphicx}
\usepackage{booktabs}
\usepackage{url}
\usepackage{caption}
\usepackage{subcaption}
\usepackage{xcolor}
\usepackage{amsmath}
\usepackage [english]{babel}
\usepackage [autostyle, english = american]{csquotes}
\MakeOuterQuote{"}
\usepackage[hidelinks]{hyperref}
\usepackage{array}
\usepackage{wrapfig}
\usepackage{multirow}
\usepackage{tabularx}
\usepackage[inline]{enumitem}
\usepackage{soul}

\newlist{inline}{enumerate*}{1}
\setlist[inline]{label=\textit{\roman*)}}

\newif\ifspace
\spacetrue

\makeindex             

\begin{document}
\title{Detection, Analysis, and Prediction of Research Topics with Scientific Knowledge Graphs}
\titlerunning{Detection, Analysis, and Prediction of Research Topics with SKGs}
%
%
\author{Angelo A. Salatino\inst{1}\orcidID{0000-0002-4763-3943} \and
Andrea Mannocci\inst{2}\orcidID{0000-0002-5193-7851} \and
Francesco Osborne\inst{1}\orcidID{0000-0001-6557-3131} 
}
\authorrunning{A. Salatino et al.}

%
\institute{Knowledge Media Institute - The Open University, \\Milton Keynes, United Kingdom\\
\email{\{angelo.salatino, francesco.osborne\}@open.ac.uk} \and
Istituto di	Scienza	e Tecnologie dell’Informazione ``A. Faedo'',\\ Italian	National Research Council, Pisa, Italy\\
\email{andrea.mannocci@isti.cnr.it}}
\maketitle              
\begin{abstract}

Analysing research trends and predicting their impact on academia and industry is crucial to gain a deeper understanding of the advances in a research field and to inform critical decisions about research funding and technology adoption.
In the last years, we saw the emergence of several publicly-available and large-scale Scientific Knowledge Graphs fostering the development of many data-driven approaches for performing quantitative analyses of research trends.
This chapter presents an innovative framework for detecting, analysing, and forecasting research topics based on a large-scale knowledge graph characterising research articles according to the research topics from the Computer Science Ontology.
We discuss the advantages of a solution based on a formal representation of topics and describe how it was applied to produce bibliometric studies and innovative tools for analysing and predicting research dynamics.
\keywords{Scholarly Communication  \and Scientific Knowledge Graphs \and Research Trends \and Research Topics \and Research Trends}
\end{abstract}


%
%
%
\section{Introduction}
Analysing research trends and predicting their impact on academia and industry is key to gain a deeper understanding of the research advancements in a field and to inform critical decisions about research funding and technology adoption.
These analyses were initially performed through qualitative approaches in a top-down fashion, relying on experts’ knowledge and manual inspection of the state of the art~\cite{kitchenham2007guidelines,wohlin2013systematic}.
However, the ever-increasing number of research publications~\cite{bornmann2015} has made this manual approach less feasible~\cite{osborne2019reducing}. 

In the last years, we witnessed the emergence of several publicly-available, large-scale Scientific Knowledge Graphs (SKGs)~\cite{hogan2020knowledge}, describing research articles and their metadata, e.g., authors, organisations, keywords, and citations. 
These resources fostered the development of many bottom-up, data-driven approaches to perform quantitative analyses of research trends.
These methods usually cover one or more of three tasks: 
\begin{inline}
  \item  detection of research topics,
  \item  scientometric analyses, and
  \item  prediction of research trends.
\end{inline}
In the first task, the articles are classified according to a set of research topics, typically using probabilistic topics models or different types of classifiers~\cite{blei2003,bolelli2009,kandimalla2020,boyack2014,zhang2018}. 
In the second task, the topics are analysed according to different bibliometrics over time~\cite{jo2007,he2009,dicaro2017}. 
For instance, each topic may be associated with the number of relevant publications or citations across the years in order to determine if their popularity in the research community is growing or decreasing.
In the third task, regression techniques or supervised forecasters are typically used for predicting the future performance of research topics~\cite{widodo2011,krampen2011,salatino2018b}.

This chapter presents an innovative framework for detecting, analysing, and forecasting research topics based on a large-scale knowledge graph characterising research articles according to the research topics from the Computer Science Ontology (CSO)~\cite{salatino2020,salatino2018}. In describing this framework we leverage and reorganise our previous research works.
Specifically, we discuss the advantage of a solution based on a formal representation of research topics and describe how it was applied to produce bibliometrics studies~\cite{kirrane2020decade,mannocci2019evolution} and novel approaches for analysing~\cite{osborne2019reducing,salatino2020b,angioni2020} and predicting~\cite{salatino2020b,salatino2018b,osborne2017ttf} research dynamics. 


The chapter is articulated in six sections. In Section~\ref{lr}, we review state-of-the-art approaches for topic detection, analysis, and forecasting. In Section~\ref{detection}, we formally define a scientific knowledge graph, then we describe CSO~\cite{salatino2020} and the CSO Classifier~\cite{salatino2019}, which is a tool for automatically annotating research papers with topics drawn from CSO. We then discuss how we generate the Academia/Industry DynAmics (AIDA) Knowledge Graph~\cite{angioni2020a}, a new SKG that describes research articles and patents according to the relevant research topics.  

Section~\ref{analysis} describes three recent methods for analysing research trends that adopt this framework:
\begin{itemize}
  \item \textit{the EDAM methodology}~\cite{osborne2019reducing} for supporting systematic reviews (Section~\ref{analysis-sr});
  \item \textit{the ResearchFlow framework}~\cite{salatino2020b} for analysing trends across academia and industry (Section~\ref{analysis-rf});
  \item  \textit{the AIDA Dashboard}~\cite{angioni2020} for supporting trend analysis in scientific conferences (Section~\ref{analysis-aida}). 
\end{itemize}

In Section~\ref{forecast}, we discuss three approaches that use the same framework for forecasting trends:
\begin{itemize}
  \item \textit{Augur}~\cite{salatino2018b} (Section~\ref{forecast-augur}), a method for predicting the emergence of novel research topics;
  \item \textit{the ResearchFlow forecaster}~\cite{salatino2020b} (Section~\ref{forecast-rf}), a supervised deep learning classifier for predicting the impact of research topics on the industrial sector;
  \item \textit{the Technology-Topic Framework (TTF)}~\cite{osborne2017ttf} (Section~\ref{forecast-techonologies}), an approach for predicting the technologies that will be adopted by a research field.
\end{itemize}

Finally, in Section~\ref{conclusions}, we outline the conclusions and future work. 

\section{Literature Review}
\label{lr}
Detecting, analysing and predicting topics in the scientific literature has attracted considerable attention in the last two decades~\cite{bolelli2009,tseng2009,decker2007,erten2004,lv2011,salatino2018classifying}. 
In this section, we review the state of the art according to the main themes of this chapter. In particular, we initially review existing approaches for detecting topics in scholarly corpora. Then, we show approaches performing the analysis of research trends and finally, we describe the approaches for predicting research trends.

\subsection{Topic detection}
Topic detection in the scholarly communication field aims at identifying relevant subjects within a set of scientific documents. This operation can support several subsequent tasks, such as suggesting relevant publications~\cite{thanapalasingam2018}, assisting the annotation of articles~\cite{peroni2017research} and video lessons~\cite{borges2019semantic}, assessing academic impact~\cite{chatzopoulos2020artsim}, recommending items~\cite{cena2013anisotropic,likavec2015property}, and building knowledge graphs of research concepts (e.g., AI-KG~\cite{dessi2020ai},  ORKG~\cite{jaradeh2019open}, TKG~\cite{rossanez2020representing}).

In the body of literature, we can find several approaches that can be characterised according to four main categories:
\begin{inline}
    \item topic modelling~\cite{blei2003,he2009,dicaro2017},
    \item supervised machine learning approaches~\cite{kandimalla2020,mai2018,garciasilva2021},
    \item approaches based on citation networks~\cite{boyack2014}, and
    \item approaches based on natural language processing~\cite{decker2007,jo2007}.
\end{inline}

For what concerns the topic modelling category, we can find the Latent Dirichlet Analysis (LDA) developed by Blei et al.~\cite{blei2003}. 
LDA is a three-level hierarchical Bayesian model to retrieve latent – or hidden – patterns in texts. 
The basic idea is that each document is modelled as a mixture of topics, where a topic is a multinomial distribution over words, which is a discrete probability distribution defining the likelihood that each word will appear in a given topic. 
In brief, LDA aims to discover the latent structure, which connects words to topics and topics to documents. 
Over the years, LDA influenced many other approaches, such as Griffiths et al.~\cite{griffiths2004} and Bolelli et al.~\cite{bolelli2009} in which they designed generative models for document collections.
Their models simultaneously modelled the content of documents and the interests of authors. Besides, Bolelli et al.~\cite{bolelli2009} exploit citations information to evaluate the main terms in documents. 
However, since topic modelling aims at representing topics as a distribution of words, it is often tricky to map them to research subjects.

For supervised machine learning approaches, Kandimalla et al.~\cite{kandimalla2020} propose a deep attentive neural network for classifying papers according to 104 Web of Science\footnote{Web of Science - \url{https://clarivate.com/webofsciencegroup/solutions/web-of-science}} (WoS) subject categories. 
Their classifier was trained on 9 million abstracts from WoS, and it can be directly applied to abstract without the need for additional information, e.g., citations. 
However, the fact that they can map research papers to a limited set of only 104 subject categories is due to the intensive human labelling effort behind the generation of a gold standard that foresees all possible research topics, and that is also balanced with regard to the number of papers labelled per topic. 
Indeed, broad areas tend to have many published papers and thus are highly represented, while very specific areas tend to have fewer papers. 

Among the citation network approaches, there is Boyack and Klavans~\cite{boyack2014}  who developed a map of science using 20 million research articles over 16 years using co-citation techniques. 
Through this map, it is possible to observe the disciplinary structure of science in which papers of the same area tend to group together. 
The main drawback of citation-based approaches is that they are able to assign each document to one topic only, while a document is seldom monothematic.

For the category of natural language processing approaches, we find Decker~\cite{decker2007} who introduced an unsupervised approach that generates paper-topic relations by exploiting keywords and words extracted from the abstracts in order to analyse the trends of topics on different timescales. 
Additionally, Duvvuru et al.~\cite{duvvuru2013} relied on keywords to build their co-occurring network and subsequently perform statistical analysis by calculating degree, strength, clustering coefficient, and end-point degree to identify clusters and associate them to research topics. Some recent approaches use word embeddings aiming to quantify semantic similarities between words. For instance, Zhang et al.~\cite{zhang2018} applied clustering techniques on a set of words represented as embeddings. 
However, all these approaches to topic detection need to generate the topics from scratch rather than exploiting a domain vocabulary, taxonomy, or ontology, resulting in noisier and less interpretable results~\cite{osborne2015}.

In brief, state-of-the-art approaches for detecting topics either use keywords as proxies for topics, or match terms to manually curated taxonomies, or apply statistical techniques to associate topics to bags of words. 

\ifspace
Unfortunately, most of these solutions are suboptimal as they 
\begin{inline}
 \item fail to manage polysemies and synonyms, respectively, when a keyword may denote different topics depending on the context and when multiple labels exist for the same research area (i.e., productive ambiguity vs maximal determinacy); and 
 \item fail to model and take advantage of the semantic relations that may hold between research areas, treating them as lists of unstructured keywords. 
\end{inline}
\fi

\subsection{Research trends analysis}
Research trends analysis deals with the dynamic component of topics, as it aims at observing their development over time. 
In the literature, we can find a variety of approaches based on citation patterns between documents~\cite{he2009,jo2007}, or co-word analysis~\cite{dicaro2017}. 
Jo et al.~\cite{jo2007} developed an approach that combines the distribution of terms (i.e., n-grams) with the distribution of the citation graph related to publications containing that term. 
They assume that if a term is relevant for a topic, documents containing that term will have a stronger connection (i.e., citations) than randomly selected ones. 
Then, the algorithm identifies the set of terms having citation patterns that exhibit synergy. Similarly, He et al.~\cite{he2009} combined the citation network with Latent Dirichlet Allocation (LDA)~\cite{blei2003}. 
They detect topics in independent subsets of a corpus and leverage citations to connect topics in different time frames. 
However, these approaches suffer from time lag, as newly published papers usually require some time, if not several years, before getting noticed and being cited~\cite{VanRaan2004}.

Di Caro et al.~\cite{dicaro2017} designed an approach for observing how topics evolve over time. After splitting the collection of documents according to different time windows, their approach selects two consecutive slices of the corpus, extracts topics using LDA, and analyses how such topics change from one time window to the other.
The central assumption is that by comparing the topics generated in two adjacent time windows, it is possible to observe how topics evolve as well as capture their birth and death. 

Other methods for analysing trends in research include overlay mapping techniques, which create maps of science and enable users to assess trends~\cite{boyack2005,leydesdorff2013}. Although these approaches provide a global perspective, the interpretation of such maps heavily relies on visual inspection by human experts.

\subsection{Research trends prediction}
Predicting research trends deals with anticipating the emergence of new topics. 
This task is particularly complex, mostly because of the limited availability of gold standards, i.e., extensive datasets of publication records manually and reliably labelled with research topics, that can be used to train a machine learning model to forecast trends. 
Indeed, regarding this aspect, we can find a very limited number of approaches in the literature.

Widodo et al.~\cite{widodo2011} performed a time series analysis to make predictions on PubMed data, and their experiments show that such time series are more suitable for machine learning techniques rather than statistical approaches. Instead, Krampen et al.~\cite{krampen2011} also rely on time series computing the Single Moving Average method to predict research trends in psychology.
Watatani et al.~\cite{watatani2013} developed a bibliometric approach based on co-citation analysis, which provides insights for technology forecasting and strategic research, as well as development planning and policy-making.

In brief, in the literature, we can find many approaches capable of tracking topics over time. However, they focus on recognised topics, which are already associated with a good number of publications. Forecasting new research topics is yet an open challenge.

\section{Topic Detection with Scientific Knowledge Graphs}
\label{detection}
In this section, we define the Scholarly Knowledge Graph (Section~\ref{skgs}), and in particular, we introduce the main one used to fuel our research: Microsoft Academic Graph (MAG).
In Section~\ref{onto-detection}, we describe the Computer Science Ontology (CSO) and the CSO Classifier, which are at the foundation of our ontology-based topic detection method.
Finally, in Section~\ref{aida}, we showcase the Academia/Industry DynAmics (AIDA) Knowledge Graph, an SKG derived by enriching MAG with the topics from CSO.

\subsection{KG of scholarly data}
\label{skgs}
Knowledge graphs are large collections of entities and relations describing real-world objects and events, or abstract concepts. Scientific Knowledge Graphs focus on the scholarly domain and describe the actors (e.g., authors, organisations), the documents (e.g., publications, patents), the research knowledge (e.g., research topics, tasks, technologies), and any other contextual information (e.g., project, funding) in an interlinked manner.
Such descriptions have formal semantics allowing both computers and people to process them efficiently and unambiguously. 
Being interconnected within a network, each entity contributes to the description of the entities related to it, providing a wider context for their interpretation.
SKGs provide substantial benefits to researchers, companies, and policymakers by powering several data-driven services for navigating, analysing, and making sense of research dynamics. Some examples include 
Microsoft Academic Graph (MAG)\footnote{Microsoft Academic Graph - \url{https://www.microsoft.com/en-us/research/project/academic}}~\cite{Wang2020}, 
AMiner~\cite{Tang2008},
ScholarlyData\footnote{ScholarlyData - \url{http://www.scholarlydata.org}}~\cite{Nuzzolese2016}, 
PID Graph\footnote{PID Graph - \url{https://www.project-freya.eu/en/pid-graph/the-pid-graph}}~\cite{fenner2019introducing}, 
SciGraph\footnote{SciGraph datasets -  \url{https://sn-scigraph.figshare.com}},
Open Research Knowledge Graph\footnote{Open Research Knowledge Graph - \url{https://www.orkg.org/orkg}}~\cite{jaradeh2019open}, 
OpenCitations\footnote{OpenCitations - \url{https://opencitations.net}}~\cite{Peroni2020}, 
and OpenAIRE research graph\footnote{OpenAIRE research graph - \url{https://graph.openaire.eu}}~\cite{Manghi2020}. 

More formally, given a set of entities E, and a set of relations R, a scientific knowledge graph is a directed multi-relational graph $G$ that comprises triples (subject, predicate, object) and is a subset of the cross product $G \subseteq E \times R \times E$.

\begin{figure}
    \center{\includegraphics[width=\textwidth]
    {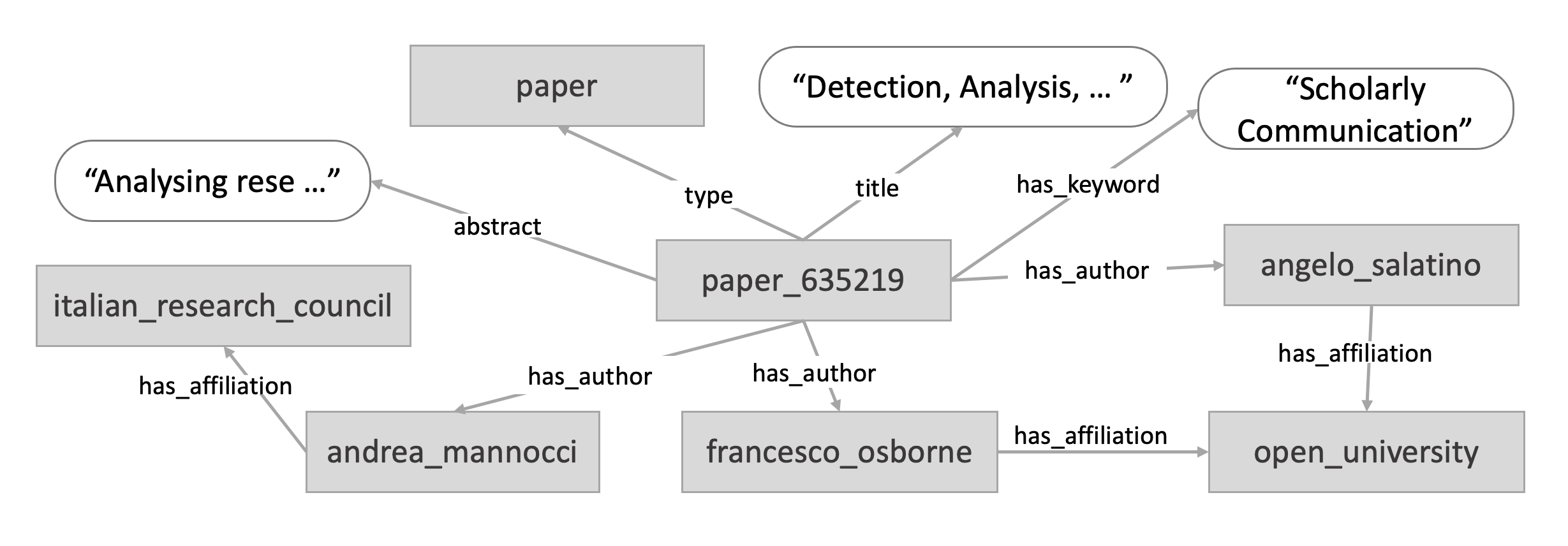}}
    \caption{\label{fig:kg} Visualisation of a scientific knowledge graph. Entities are represented as rectangles and relations as arrows.}
\end{figure}

Figure~\ref{fig:kg} depicts an excerpt of a scientific knowledge graph representing some metadata about this current chapter.  

In this chapter, we will focus on \textit{Microsoft Academic Graph} (MAG)~\cite{Wang2020,sinha2015}, a pan-publisher, longitudinal scholarly knowledge graph produced and actively maintained by Microsoft, which contains scientific publication records, citation relations, authors, institutions, journals, conferences, and fields of study.

We decided to adopt MAG because it is the most extensive datasets of scholarly data publicly available~\cite{visser2021}, containing more than 250 million publications as of January 2021. In addition, it is widely adopted by the scientometrics community, making our results easier to replicate.

At the time of writing, MAG is released at no cost (\textit{per se}) for research purposes, while a small fee is required to cover the costs and maintenance of the provisioning infrastructure, which is based on Azure.
MAG can be either processed on the cloud, leveraging Azure cloud computing solutions (whose usage and costs fall outside the scope of the present chapter) or downloaded as a whole.
In the latter case, the current due costs are around 100\$ a go for bandwidth (likely to increase slightly over time as the dataset grows and is regenerated fortnightly), plus a few dollars a month for data parking in case the dataset is kept on the cloud as a backup, rather than deleted right away.
In this form, MAG is provided as a collection of TSV files via Microsoft Azure Storage and features an ODC-BY\footnote{Open Data Commons Attribution Licence (ODC-BY) v1.0 - \url{https://opendatacommons.org/licenses/by/1.0}} licence, an aspect that is essential to ensure the transparency and reproducibility of any resulting analysis in compliance with Open Science best practices.
The information contained in this dataset consists of six types of main entities in the scholarly domain — publications, authors, affiliations, venues (journals and conferences), fields of study, and events (specific conference instances).
These entities are connected through relations like citations, authorship, and others. 
The relations between such entities are described in details in~\cite{sinha2015}.
The dataset contains publication metadata, such as DOI, title, year of publication, and venue of publication; moreover, MAG provides, whenever possible, key data often not available in other publicly accessible datasets (e.g., Crossref\footnote{Crossref API -  \url{https://github.com/CrossRef/rest-api-doc}}, DBLP\footnote{DBLP - \url{https://dblp.uni-trier.de}}), including authors' and affiliations' identifiers, which are often required to address some compelling research questions (e.g., collaboration networks).
Although paper abstracts are available for many records in MAG, this, however, does not redistribute the publications' full-texts, somehow limiting full-scale NLP applications.

For the sake of completeness, in the case current costs and technical skills associated with MAG provisioning are not viable, other alternative SKGs repackaging MAG are available for free.
\textit{The Open Academic Graph}\footnote{Open Academic Graph - \url{https://www.microsoft.com/en-us/research/project/open-academic-graph}} (OAG) is a large SKG unifying two billion-scale academic graphs: MAG and AMiner. 
In mid-2017, the first version of OAG, which contains 166,192,182 papers from MAG and 154,771,162 papers from AMiner, and 64,639,608 linking (matching) relations between the two graphs, was released. 
The current version of OAG combines the MAG snapshot as of November 2018 and AMiner snapshots of July 2018 and January 2019. In this new release, authors, venues, newer publication data, and the corresponding matchings are available as well.
Another feasible alternative is the \textit{OpenAIRE dataset DOIboost}~\cite{LaBruzzo2018} which provides an ``enhanced'' version of Crossref that integrates information from Unpaywall, ORCID and MAG, such as author identifiers, affiliations, organisation identifiers, and abstracts. DOIboost is periodically released on Zenodo\footnote{DOIboost laster release - \url{https://zenodo.org/record/3559699}}.

\subsection{Ontology-based topic detection}
\label{onto-detection}
In the following section, we describe the two main technologies enabling our methodology for ontology-based topic detection: the Computer Science Ontology (Section~\ref{cso}) and the CSO Classifier (Section~\ref{cso-c}).

\subsubsection{The Computer Science Ontology}
\label{cso}
The Computer Science Ontology is a large-scale ontology of research areas in the field of Computer Science. It was automatically generated using the Klink-2 algorithm~\cite{osborne2015} on a dataset of 16 million publications, mainly in the field of Computer Science~\cite{osborne2013}. 
Compared to other solutions available in the state of the art (e.g., the ACM Computing Classification System), the Computer Science Ontology includes a much higher number of research topics, which can support a more granular representation of the content of research papers, and it can be easily updated by rerunning Klink-2 on more recent corpora of publications.  

The current version of CSO\footnote{CSO is available for download at   \url{https://w3id.org/cso/downloads}} includes 14 thousand semantic topics and 159 thousand relations. The main root is Computer Science; however, the ontology includes also a few additional roots, such as Linguistics, Geometry, Semantics, and others. The CSO data model\footnote{CSO Data Model -   \url{https://cso.kmi.open.ac.uk/schema/cso}} is an extension of SKOS\footnote{SKOS Simple Knowledge Organisation System -   \url{http://www.w3.org/2004/02/skos}} and it includes four main semantic relations:  

\begin{itemize}
\item
  \texttt{superTopicOf}, which indicates that a topic is a super-area of another one (e.g., Semantic Web is a super-area of Linked Data).
\item
  \texttt{relatedEquivalent}, which indicates that two given topics can be treated as equivalent for exploring research data (e.g., Ontology Matching and Ontology Mapping).
\item
  \texttt{contributesTo}, which indicates that the research output of one topic contributes to another.
\item
  \texttt{owl:sameAs}, this relation indicates that a research concept is identical to an external resource, i.e., DBpedia.
\end{itemize}

The Computer Science Ontology is available through the CSO Portal\footnote{Computer Science Ontology Portal -   \url{https://cso.kmi.open.ac.uk}}, a web application that enables users to download, explore, and visualise sections of the ontology. Moreover, users can use the portal to provide granular feedback at different levels, such as rating topics and relations, and suggesting missing relations.

CSO has been adopted by Springer Nature, one of the two main academic publishers, that used it to support several innovative applications, including the Smart Topic Miner \cite{osborne2016}, a tool designed to assist the Springer Nature editorial team in classifying proceedings\cite{salatino2019a}, and  Smart Book Recommender \cite{thanapalasingam2018}, an ontology-based recommend system for scientific volumes.

In the last few years CSO supported the creation of many innovative approaches and applications, including ontology-driven topic models (e.g., CoCoNoW~\cite{beck2020automatic}), recommender systems for articles (e.g., SBR~\cite{thanapalasingam2018}) and video lessons~\cite{borges2019semantic},
tools for detecting research communities
(e.g., TST~\cite{osborne2014},  RCMB~\cite{osborne2014hybrid}),
visualisation frameworks (e.g., ScholarLensViz~\cite{ScholarLensViz2020}, ConceptScope~\cite{zhang2021conceptscope}), temporal knowledge graphs (e.g., TGK~\cite{rossanez2020representing}), NLP frameworks for entity extraction~\cite{dessi2021generating},
 knowledge graph embeddings (e.g., Trans4E~\cite{nayyeri2021trans4e}),
tools for identify domain experts (e.g., VeTo~\cite{vergoulis2020veto}),
and  systems for predicting academic impact (e.g., ArtSim~\cite{chatzopoulos2020artsim}), 
research topics (e.g., Augur~\cite{salatino2018b}),
ontology concepts (e.g., SIM~\cite{cano2016}, POE~\cite{osborne2018pragmatic}),
and technologies (e.g., TTF~\cite{osborne2017ttf}, TechMiner~\cite{osborne2016techminer}).
It was also used for several large-scale analyses of the literature (e.g., Cloud Computing~\cite{lulaadvanced}, Software Engineering~\cite{chicaiza2020using}, Ecuadorian publications~\cite{chicaiza2020using}).

\subsubsection{The CSO Classifier}
\label{cso-c}
\begin{figure}[!htb]
\center{\includegraphics[width=\textwidth]
{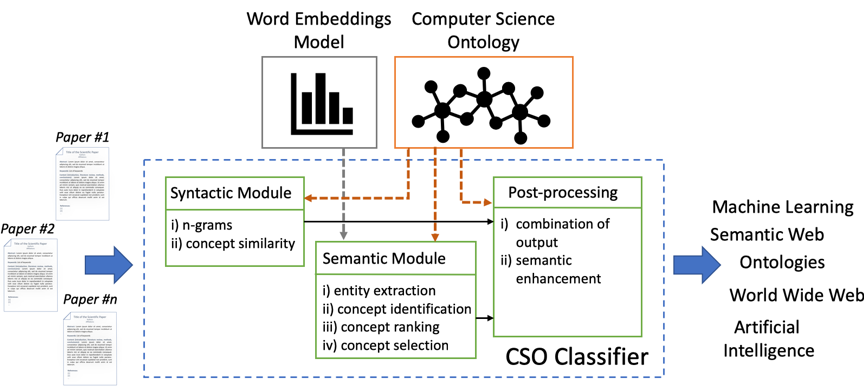}}
\caption{\label{fig:cso-workflow} Workflow of the CSO Classifier.}
\end{figure}

The CSO Classifier~\cite{salatino2019} is a tool for automatically classifying research papers according to the Computer Science Ontology. 
This application takes in input the metadata associated with a research document (title, abstract, and keywords) and returns a selection of research concepts drawn from CSO, as shown in Figure~\ref{fig:cso-workflow}.

The CSO Classifier works in three steps. First, it finds all topics in the ontology that are explicitly mentioned in the paper (\textit{syntactic module}). 
Then it identifies further semantically related topics by means of part-of-speech tagging and word embeddings (\textit{semantic module}). Finally, it enriches this set by including the super-areas of these topics according to CSO (\textit{post-processing module}). 

In particular, the syntactic module splits the text into unigrams, bigrams, and trigrams. Each n-gram is then compared with concept labels in CSO using the Levenshtein similarity. 
As a result, this module returns all matched topics having similarities greater than or equal to the pre-defined threshold.

The semantic module instead leverages a pre-trained Word2Vec word embedding model that captures the semantic properties of words~\cite{mikolov2013}. We trained this model\footnote{The model parameters are: \emph{method} = skipgram, \emph{embedding-size} = 128, \emph{window-size} = 10, \emph{min-count-cutoff} = 10, \emph{max-iterations} = 5.} using titles and abstracts of over 4.6 million English publications in the field of Computer Science from MAG. 
We preprocessed this data by replacing spaces with underscores in all n-grams matching the CSO topic labels (e.g., “semantic web” turned into “semantic\_web”). 
We also performed a collocation analysis to identify frequent bigrams and trigrams (e.g., “highest\_accuracies”, “highly\_cited\_journals”)~\cite{mikolov2013}. This solution allows the classifier to disambiguate concepts better and treat terms such as “deep\_learning” and “e-learning” as entirely different words. 

More precisely, to compute the semantic similarity between the terms in the document and the CSO concepts, the semantic module uses part-of-speech tagging to identify candidate terms composed by a combination of nouns and adjectives and decomposes them into unigrams, bigrams, and trigrams. For each n-gram, it retrieves its most similar words from the Word2Vec model. The n-gram tokens are initially glued with an underscore, creating one single word, e.g., “semantic\_web”. If this word is not available within the model vocabulary, the classifier uses the average of the embedding vectors of all its tokens. Then, it computes the relevance score for each topic in the ontology as the product between the number of times it was identified in those n-grams (frequency) and the number of unique n-grams that led to it (diversity). Finally, it uses the elbow method~\cite{satopaa2011} for selecting the set of most relevant topics. 

The resulting set of topics returned by the two modules is enriched in the post-processing module by including all their super-topics in CSO. For instance, a paper tagged as \textit{neural network} is also tagged with \textit{machine learning} and \textit{artificial intelligence}. This solution yields an improved characterisation of high-level topics that are not directly referred to in the documents. More details about the CSO Classifier are available in~\cite{salatino2019}.

The latest release of the CSO Classifier can be installed via \textit{pip} from PyPI with \texttt{pip install cso-classifier}; or it can be simply downloaded from \url{https://github.com/angelosalatino/cso-classifier}.

\subsection{The Academia/Industry DynAmics Knowledge Graph}
\label{aida}
In order to support new approaches for monitoring and predicting research trends across academia and industry, we build on CSO and the CSO Classifier to generate the Academia/Industry DynAmics (AIDA) Knowledge Graph~\cite{angioni2020a}. This novel resource describes 21 million publications and 8 million patents according to the research topics drawn from CSO. 5.1 million publications and 5.6 million patents are further characterised according to the type of the author's affiliation (e.g., academia, industry) and 66 industrial sectors (e.g., automotive, financial, energy, electronics) organised in a two-level taxonomy. 

AIDA was generated using an automatic pipeline that integrates several knowledge graphs and bibliographic corpora, including Microsoft Academic Graph, Dimensions, English DBpedia, the Computer Science Ontology~\cite{salatino2018}, and the Global Research Identifier Database (GRID). 
Its pipeline consists of three phases: 
\begin{inline}
  \item topics extraction,
  \item extraction of affiliation types, and
  \item industrial sector classification. 
\end{inline}

To extract topics from publications and patents, we run the CSO Classifier.
Instead, to extract the affiliation types of a given document, we queried GRID\footnote{Global Research Identifier Database (GRID)
 - \url{https://grid.ac}}, an open database identifying over 90 thousand organisations involved in research. 
GRID describes research institutions with a multitude of attributes, including their type, e.g., education, company, government, and others. 
A document is classified as ``academic'' if all its authors have an educational affiliation and as ``industrial'' if all its authors have an industrial affiliation. 
Documents whose authors are from both academia and industry are classified as ``collaborative''.

Finally, we characterised documents from industry according to the Industrial Sectors Ontology (INDUSO)\footnote{INDUSO - \url{http://aida.kmi.open.ac.uk/downloads/induso.ttl}}. 
In particular, from GRID, we retrieved the Wikipedia entry of companies, and we queried DBpedia, the knowledge graph of Wikipedia. 
Then, we extracted the objects of the properties ``\texttt{About:Purpose}'' and ``\texttt{About:Industry}'', and mapped them to INDUSO sectors. 
All industrial documents were associated with the industrial sectors based on their affiliations.
\begin{table}[ht] 
\centering
\caption{Distribution of publications and patents classified as Academia, Industry and Collaboration.\label{tab:overview}}
\begin{tabular}{l|r|r}
\toprule
\multicolumn{1}{l|}{\textbf{}} & \multicolumn{1}{l|}{\textbf{Publications}} & \multicolumn{1}{l}{\textbf{Patents}} \\ \midrule
\textit{Total documents}                 & 
20,850,710                                 & 7,940,034          \\
\textit{Documents with GRID IDs}                 & 5,133,171                                  & 5,639,252                            \\ \midrule
\textit{Academia}            & 3,585,060                                  & 133,604                              \\
\textit{Industry}              & 787,151                                    & 4,741,695                            \\
\textit{Collaborative}         & 133,781                                    & 16,335                               \\
\textit{Additional categories with GRID ID} & 627,179 & 747,618  \\
\bottomrule
\end{tabular}
\end{table}

Table~\ref{tab:overview} reports the number of publications and patents from academy, industry, and collaborative efforts. Most scientific publications (69.8\%) are written by academic institutions, but industry is also a strong contributor (15.3\%). Conversely, 84\% of the patents are from industry and only 2.3\% from academia. 
Collaborative efforts appear limited, including only 2.6\% of the publications and 0.2\% of patents.

The data model of AIDA is available at \url{http://aida.kmi.open.ac.uk/ontology} and it builds on SKOS\footnote{SKOS Simple Knowledge Organisation System -   \url{http://www.w3.org/2004/02/skos}} and CSO\footnote{CSO Data Model - \url{https://cso.kmi.open.ac.uk/schema/cso}}.
It focuses on four types of entities: \textit{publications, patents, topics,} and \textit{industrial sectors}.

\ifspace
The main information about publications and patents are given by means of the following semantic relations:
\begin{itemize}
  \item \texttt{hasTopic}, which associates to the documents all their relevant topics drawn from CSO;
  \item \texttt{hasAffiliationType} and \texttt{hasAssigneeType}, which associates to the documents the three categories (academia, industry, or collaborative) describing the affiliations of their authors (for publications) or assignees (for patents);
  \item \texttt{hasIndustrialSector}, which associates to documents and affiliations the relevant industrial sectors drawn from the Industrial Sectors Ontology (INDUSO) we describe in the next sub-section.
\end{itemize}
\fi

A dump of AIDA in Terse RDF Triple Language (Turtle) is available at \url{http://aida.kmi.open.ac.uk/downloads}.

\section{Research Trends Analysis}
\label{analysis}
In this section, we discuss different methodologies that use the framework described in the previous section for analysing the evolution of research topics in time. 
In particular, we first introduce the EDAM methodology for supporting systematic reviews (Section~\ref{analysis-sr}) and present two exemplary bibliometric analyses, then we describe the ResearchFlow approach for analysing research trends across academia and industry  (Section~\ref{analysis-rf}), and finally, we introduce the AIDA Dashboard, a web application for examining research trends in scientific conferences (Section~\ref{analysis-aida}).

\subsection{The EDAM Methodology for Systematic Reviews} \label{analysis-sr}

A systematic review is {\em ``a means of evaluating and interpreting all available research relevant to a particular research or topic area or phenomenon of interest''}~\cite{kitchenhamTR2004}. 
Systematic reviews are used by researchers to create a picture of the state of the art in a research field, often including the most important topics and their trends.
Given a set of research questions, and by following a systematically defined and reproducible process, a systematic review can assist the selection of a set of research articles (primary studies) that contribute to answering them~\cite{petersen2015guidelines}. 

In this section, we describe the EDAM (Expert-Driven Automatic Methodology)~\cite{osborne2019reducing}, which is a recent methodology for creating systematic reviews, thus limiting the number of tedious tasks that have to be performed by human experts.  Specifically, EDAM can produce a comprehensive analysis of topic trends in a field by 
\begin{inline}
  \item characterising the area of interest using an ontology of topics,
  \item asking domain experts to refine this ontology, and 
  \item exploiting this knowledge base for classifying relevant papers and producing useful analytics.
\end{inline}
The main advantage of EDAM is that the authors of the systematic review are still in full control of the ontological schema used to represent the topics. 
\begin{figure}[htpb]
\centering
\includegraphics[width=.9\textwidth]{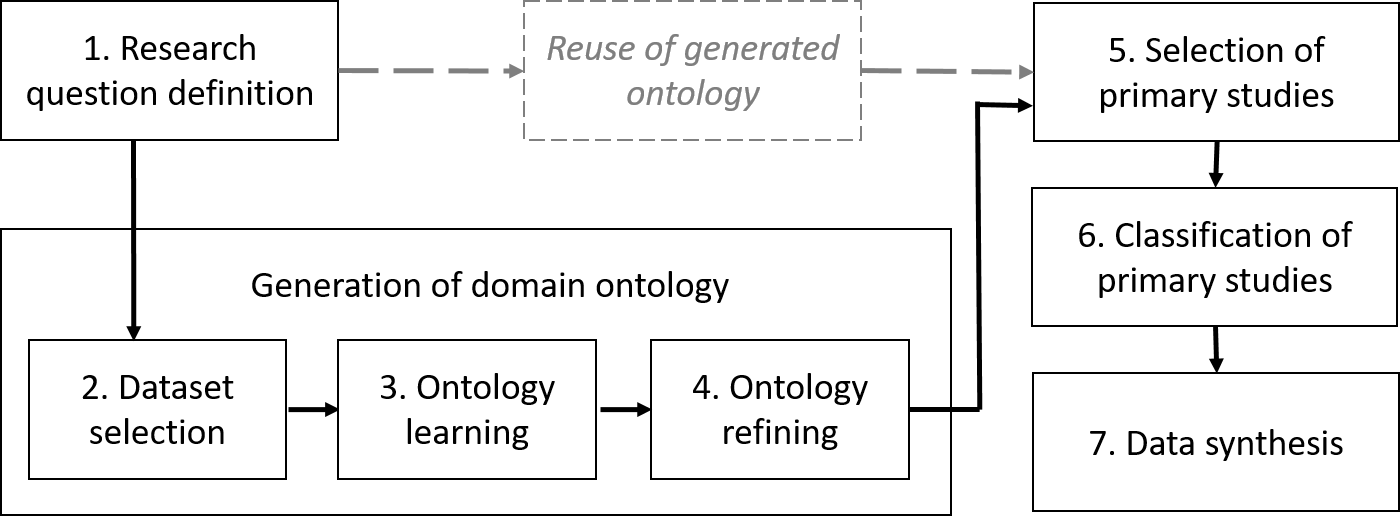}
\caption{Steps of a systematic mappings adopting the EDAM methodology. The gray-shaded elements refer to the alternative step of reusing the previously generated ontology.}\label{fig:edam}
\end{figure}	

Figure~\ref{fig:edam} illustrates the workflow of EDAM.
First, the researchers define the research questions that will be answered by the produced analytics and select the data sources. 
Then, they reuse or create the ontology that will support the classification process. 
The creation can also be supported by automatic methodologies for ontology learning such as
\begin{inline}
  \item statistical methods for deriving taxonomies from keywords~\cite{liu2012automatic};
  \item natural language processing approaches (e.g., FRED~\cite{gangemi2017semantic}, LODifier~\cite{augenstein2012lodifier}, Text2Onto~\cite{cimiano2005text2onto});
  \item approaches based on deep learning (e.g., recurrent neural networks~\cite{petrucci2016ontology});
  \item specific approaches for generating research topic ontologies (e.g., Klink-2~\cite{osborne2015}).
\end{inline}
The resulting ontology is corrected and refined by domain experts. When the ontology is ready, the authors define the primary studies' inclusion criteria according to the domain ontology and other metadata of the papers (e.g., year, venue, language).
The inclusion criteria are typically expressed as a search string, which uses simple logic constructs, such as  AND, OR, and NOT~\cite{aromataris2014constructing}. The simplest way for mapping categories to papers is to associate to each category each paper that contains the label of the category or any of its subcategories. However, this step can also take advantage of a more advanced classification approach, such as the CSO Classifier described in Section~\ref{cso-c}. 
The last step is the computation of analytics to answer the research questions. 

\begin{figure}[!t]
  \centering
  \includegraphics[width=.9\textwidth]{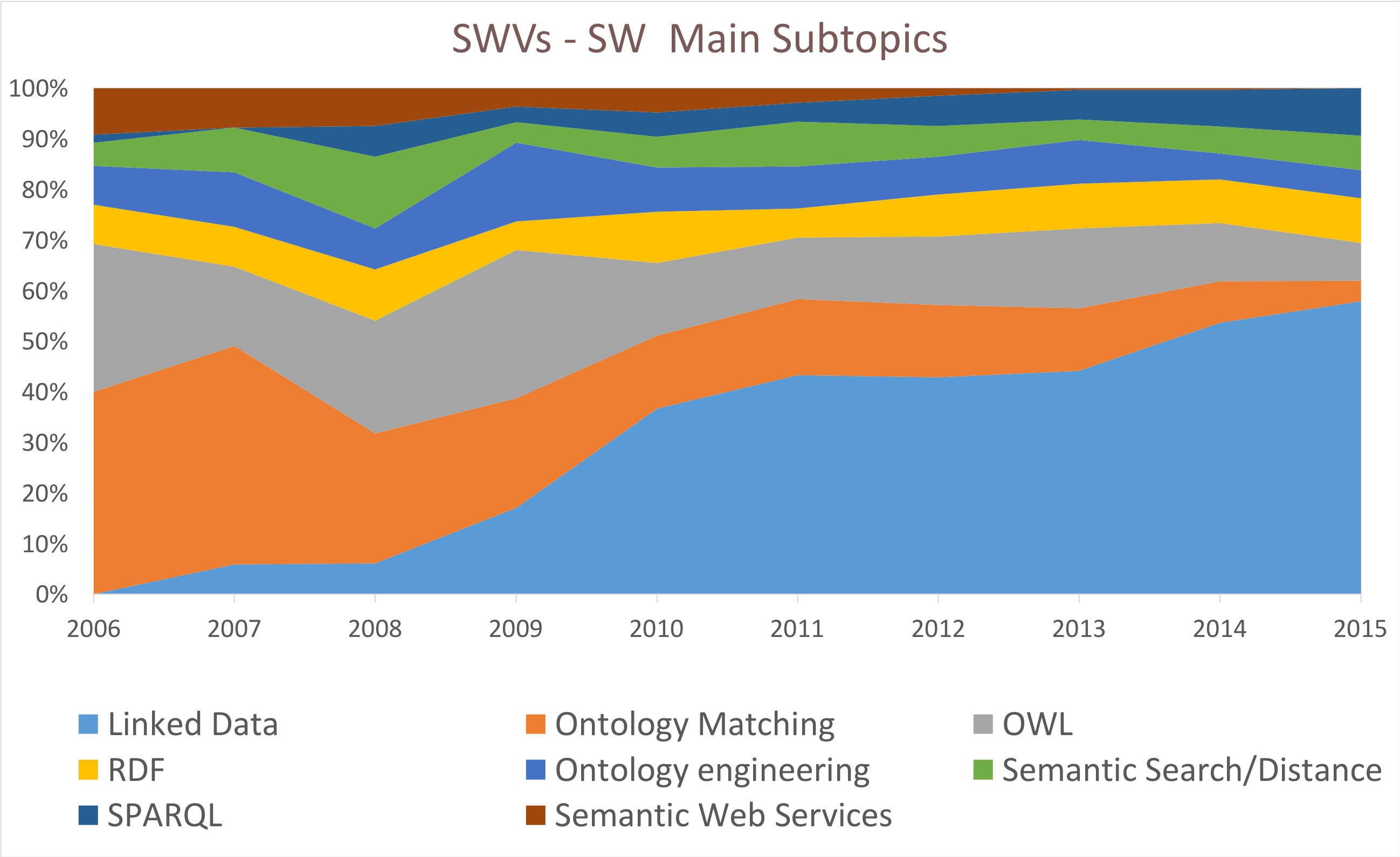}
  \caption{Exemplary analysis in Semantic Web: Percentage of articles associated with the eight main subtopics of Semantic Web.}
  \label{fig:rexplore-mv-publications-subtopics}
\end{figure}

EDAM has also been used in a recent review that analyses the main research trends in the field of Semantic Web~\cite{kirrane2020decade}. 
This study compares two datasets covering respectively the main Semantic Web venues (e.g., ISWC, ESWC, SEMANTiCS, SWJ, and JWS) and 32,431 publications associated with the Semantic Web from a dump of Scopus.
The authors annotated the articles with CSO and produced several analyses based on the number of papers and citations in each research topic over the years. 
Specifically, articles were associated with a given topic if their title, abstract, or keywords field contained:
\begin{inline}
    \item the label of the topic (e.g., "semantic web"), 
    \item a \texttt{relevantEquivalent} of the topic (e.g., "semantic web tecnologies"), 
    \item a \texttt{skos:broaderGeneric} of the topic (e.g., "ontology matching"), or 
    \item a \texttt{relevantEquivalent} of any \texttt{skos:broaderGeneric} of the topic (e.g., "ontology mapping").
\end{inline}
Figure~\ref{fig:rexplore-mv-publications-subtopics} shows, for example, the popularity of the main topics in the main venue of the Semantic Web community. 
The analysis highlights two main dynamics confirmed by domain experts: the fading of Semantic Web Services and the rapid growth of Linked Data after the release of DBpedia in 2007~\cite{auer2007dbpedia}.

A further instantiation of the EDAM methodology can be found in~\cite{mannocci2019evolution}, where Mannocci et al. performed a circumstantial, narrow-aimed study for the 50th anniversary of the International Journal of Human-Computer Studies (IJHCS) and delivered an analysis on the evolution throughout the last 50 years of the journal making an extensive comparison with another top-tier venue in the HCI research field: the CHI conference.

Among several other analyses (e.g., bibliometrics, spatial scientometrics), one aspect of the study focuses on topic analysis in both venues. To this end, it instantiates EDAM using CSO and the CSO classifier as reference ontology and classifier, respectively.

\begin{figure*}[!t]
  \begin{center}
      \includegraphics[width=.8\linewidth]{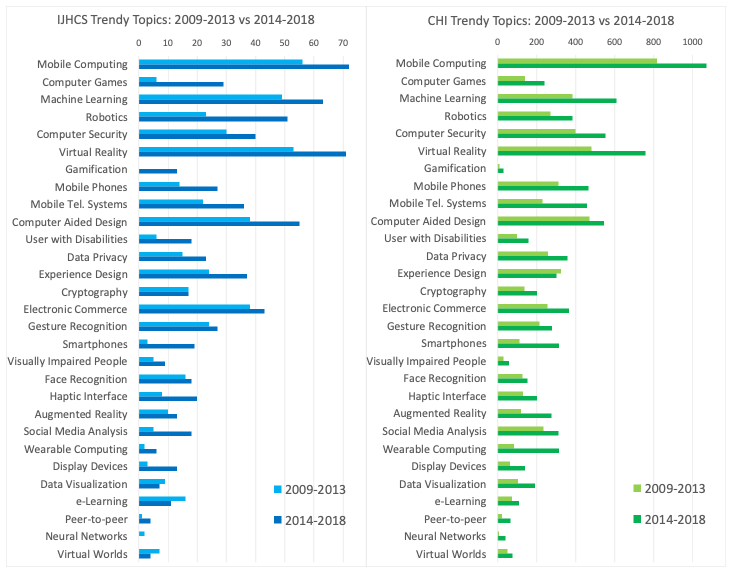}
      \caption{Exemplary analysis in Human Computer Interaction: Growing Topics in IJHCS and CHI.}
      \label{fig:ijhcs+chi_fingerprint}
  \end{center}
\end{figure*}

Figure~\ref{fig:ijhcs+chi_fingerprint} shows a comparison of the most growing topics in the two venues according to their number of articles in 2009-2013 and 2014-2018. These plots highlight very clearly which topics are gaining momentum in a community.

%

\ifspace
Figure~\ref{fig:ijhcs+chi_countries} shows the geographical distribution of some of the most important topics in the two venues under analysis. 
This kind of analytics allows to easily assess the contribution of different countries for each topic. 
For instance, Japan is ranked seventh when considering the full CHI dataset, but in Robotics is third. 
Similarly, South Korea, ranked 12th and 10th in IJHCS and CHI, is particularly active in Computer Security, ranking in fourth and sixth place when considering this topic.

\begin{figure*}[!t]
  \begin{center}
      \includegraphics[width=.9\linewidth]{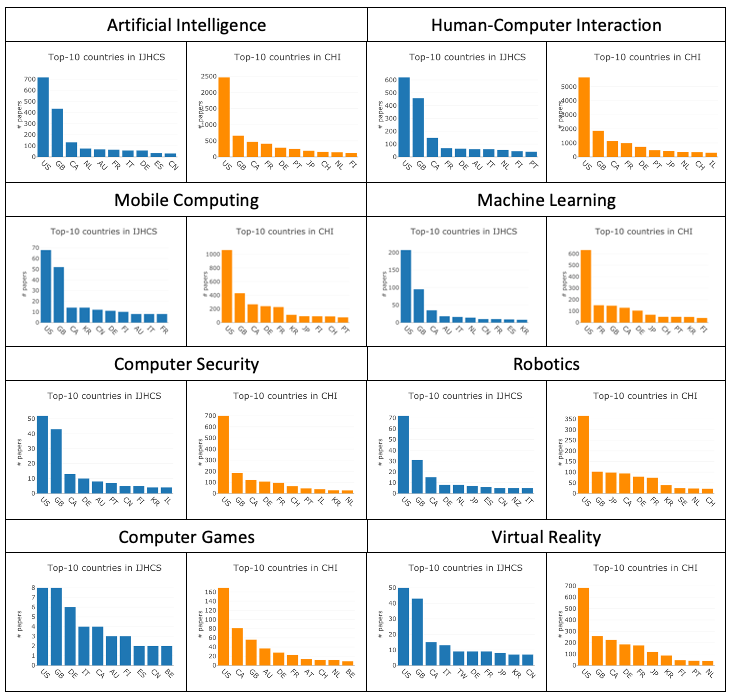}
      \caption{Overall country contributions in IJHCS and CHI across the top-8 topics.}
      \label{fig:ijhcs+chi_countries}
  \end{center}
\end{figure*}
\fi

\subsection{ResearchFlow: analysing trends across academia and industry}
\label{analysis-rf}
ResearchFlow~\cite{salatino2020b} is a recent methodology for quantifying the diachronic behaviour of research topics in academia and industry that builds on the Academia/\\Industry DynAmics (AIDA) Knowledge Graph~\cite{angioni2020a}, described in Section~\ref{aida}.

Specifically, ResearchFlow represents the evolution of research topics in CSO according to four time series reporting the frequency of 
\begin{inline}
  \item papers from academia, 
  \item papers from industry, 
  \item patents from academia, and
  \item patents from industry.
\end{inline}
Such characterisation allows us to 
\begin{inline}
  \item study the diachronic behaviour of topics in order to characterise their trajectory across academia and industry, 
  \item compare each pair of signals to understand which one typically precedes the other and in which order they usually tackle a research topic, and 
  \item assess how signals influence each other by identifying pairs of signals that are highly correlated, after compensating for a time delay. 
\end{inline}

\begin{figure}
\center{\includegraphics[width=0.8\textwidth]
{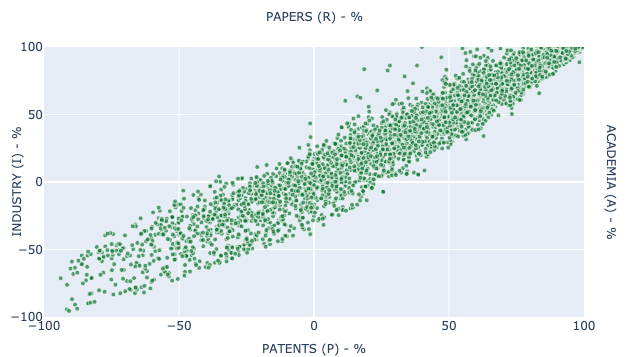}}
\caption{\label{fig:td} Distribution of the most frequent 5,000 topics according to their academia-industry and publications-papers indexes.}
\end{figure}

Figure~\ref{fig:td} shows the distribution of the most frequent 5,000 topics from CSO in a bi-dimensional diagram according to two indexes: academia-industry (x-axis) and papers-patents (y-axis).
The academia-industry index for a given topic $t$ is the difference between the documents in academia and industry related to $t$, over the whole set of documents. 
A topic is predominantly academic if this index is positive; otherwise, it is mostly populated by industry. 
On the other hand, the papers-patents index of a given topic $t$ is the difference between the number of research papers and patents related to $t$, over the whole set of documents. If this index is positive, a topic tends to be associated with a higher number of publications; otherwise, it has a higher number of patents. 

From the figure, we can observe that topics are tightly distributed around the bisector: the ones which attract more interest from academia are prevalently associated with publications (top-right quadrant). In contrast, the ones in industry are mostly associated with patents (bottom-left quadrant). 

We also analysed which time series typically precedes another in first addressing a research topics. In particular, we determined when a topic emerges in all associated signals, and then we compared the time elapsed between each couple of signals. To avoid false positives, we considered a topic as ''emerged'' when associated with at least ten documents. 
Our results showed that 89.8\% of the topics first emerged in academic publications, 3.0\% in industrial publications, 7.2\% in industrial patents, and none in academic patents. On average, publications from academia preceded publications from industry by 5.6$\pm$5.6 years, and in turn, the latter preceded patents from industry by 1.0$\pm$5.8 years. Publications from academia also preceded patents from industry by 6.7$\pm$7.4 years.

\subsection{The AIDA dashboard}\label{analysis-aida}
The framework discussed in this chapter can also be used to support tools that allow users to analyse research dynamics. This is the case of the AIDA Dashboard~\cite{angioni2020}, a novel web application for analysing scientific conferences which integrates statistical analysis, semantic technologies, and visual analytics. 
This application has been developed to support the editorial team at Springer Nature in assessing and evaluating the evolution of conferences through a number of bibliometric analytics.
The AIDA Dashboard builds on AIDA SKG and, compared to other state-of-the-art solutions, it introduces three novel features.
First, it associates conferences with a very granular representation of their topics from the CSO~\cite{salatino2018} and uses it to produce several analyses about its research trends over time.
Second, it enables an easy comparation and raking of conferences according to several metrics within specific fields (e.g., Digital Libraries) and timeframes (e.g., last five years).
Finally, it offers several parameters for assessing the involvement of industry in a conference. 
This includes the ability to focus on commercial organisations and their performance, to report the ratio of publications and citations from academia, industry,  and collaborative efforts, and to distinguish industrial contributions according to the 66 industrial sectors from INDUSO.

AIDA Dashboard is highly scalable and allows users to browse the different facets of a conference according to eight tabs: \textit{Overview}, \textit{Citation Analysis}, \textit{Organisations}, \textit{Countries}, \textit{Authors}, \textit{Topics}, \textit{Related Conferences}, and \textit{Industry}.

\begin{figure}[!htb]
\center{\includegraphics[width=\textwidth]
{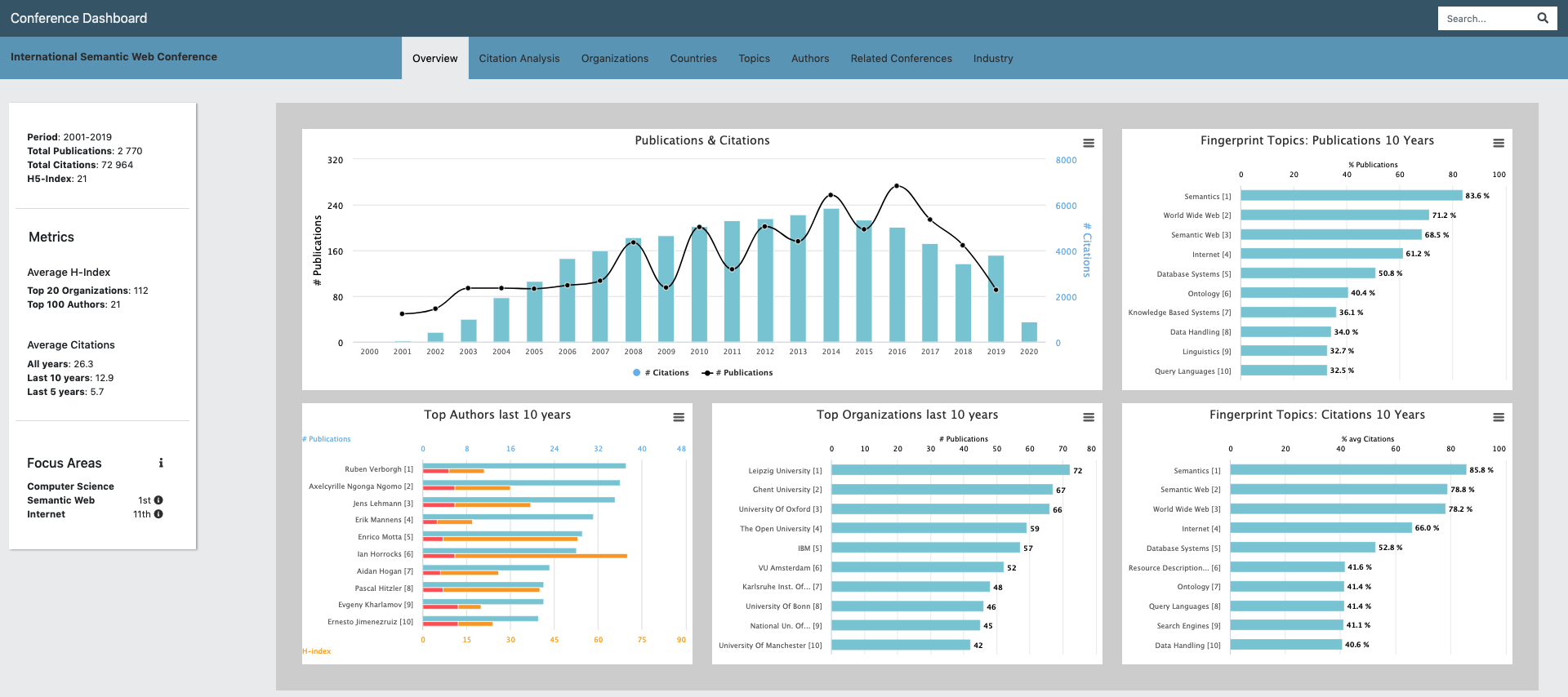}}
\caption{\label{fig:dashboard} The Overview of the International Semantic Web Conference (ISWC) according to the AIDA Dashboard.} 
\end{figure}

Figure~\ref{fig:dashboard} shows the \textit{Overview} tab of the International Semantic Web Conference (ISWC). This is the main view of a conference that provides introductory information about its performance, the main authors and organisation, and the conference rank in its main fields in terms of average citations for papers during the last five years.
More detailed and refined analytics are available in the other tabs.

\section{Research Trends Prediction}
\label{forecast}
In this section, we explore two approaches that use our framework for predicting research trends. 
In Section~\ref{forecast-augur}, we present Augur, a method for forecasting the emergence of new research trends which takes advantage of the collaborative networks of topics. 
In Section~\ref{forecast-rf}, we present an approach that was developed within the ResearchFlow framework and aims at predicting the impact of research topics on the industrial landscape.

In Section~\ref{forecast-techonologies}, we introduce the Technology-Topic Framework (TTF)~\cite{osborne2017ttf}, a novel methodology that takes advantage of a semantically enhanced technology-topic model for predicting the technologies that will be adopted by a research community.

\subsection{Predicting the emergence of research topics}
\label{forecast-augur}
Augur~\cite{salatino2018b} is an approach that aims to effectively detect the emergence of new research topics by analysing topic networks and identifying research areas exhibiting an overall increase in the pace of collaboration between already existing topics. 
This dynamic has been shown to be strongly associated with the emergence of new topics in a previous study by Salatino et al.~\cite{salatino2017}.

A topic network is a fully weighted graph in which nodes represent topics, and the edges identify whether the connected pair of topics co-occurred together in at least one research paper. Additionally, the node weights count the number of papers in which a topic appears, and edge weights count the number of papers the two topics co-appear together. The original implementation of Augur extracted these networks using CSO on the Rexplore dataset~\cite{osborne2013}. 

\begin{figure}
\center{\includegraphics[width=0.8\textwidth]
{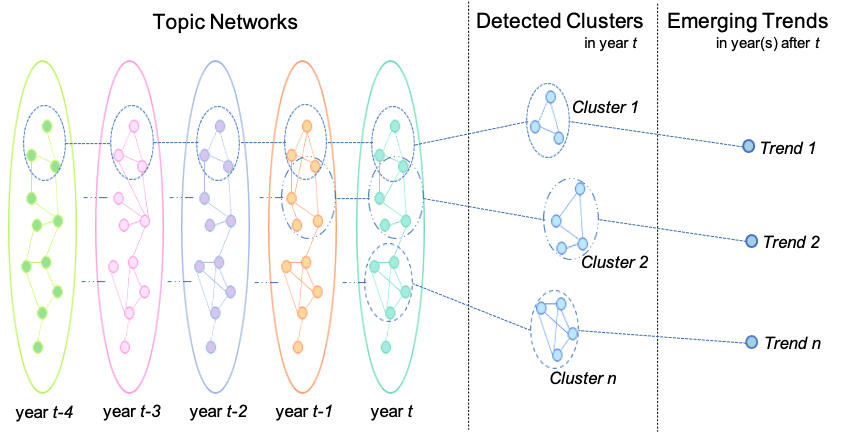}}
\caption{\label{fig:concept_augur} Augur approach. On the left, the topic networks, whereas on the right there are emerging trends. The dashed circles represent groups of research topics nurturing new research trends.}
\end{figure}

Augur takes as input five topic networks characterising the collaborations between topics over five subsequent years (e.g., 2017-2021), as shown in the left side of Fig.~\ref{fig:concept_augur}.
Over these networks, it then applies the Advanced Clique Percolation Method, a clustering algorithm for identifying groups of topics that over time intensify their pace of collaboration (centre of Fig.~\ref{fig:concept_augur}). 
The resulting clusters outline the portions of the network that may be nurturing research topics that will emerge in the following years (right side of Fig.~\ref{fig:concept_augur}).

In order to evaluate this approach, we built a gold standard of 1408 emerging trends in the 2000-2011 timeframe. Each emerging topic has been associated with a list of related topics considered as “ancestors”. The latter is crucial because Augur returns clusters of topics, which can also be considered ancestors of the future emerging trends. 
To this end, the evaluation consisted of checking whether the clusters of topics returned by Augur could be mapped with the ancestors of the emerging trends in the gold standard.

\begin{table}[h!]
\caption{Values of Precision and Recall for the five approaches along time. In bold the best results. The table and the experiments were previously reported in \cite{salatino2018b}.}\label{tab:augur-per}
\centering
\begin{tabular}{l|ll|ll|ll|ll|ll}
\toprule
\multicolumn{1}{l|}{} & \multicolumn{2}{c|}{FG} & \multicolumn{2}{c|}{LE} & \multicolumn{2}{c|}{FCM} & \multicolumn{2}{c|}{CPM} & \multicolumn{2}{c}{ACPM}  \\ \midrule
Years & Precis.         & Recall        & Precis.         & Recall        & Precis.         & Recall         & Precis.         & Recall         & Precis.          & Recall         \\\midrule
1999  & 0.27        & 0.11       & 0.00        & 0.00       & 0.00        & 0.00        & 0.06        & 0.01        & \textbf{0.86} & \textbf{0.76} \\
2000  & 0.21        & 0.07       & 0.14        & 0.02       & 0.96        & 0.01        & 0.05        & 0.00        & \textbf{0.78} & \textbf{0.70} \\
2001  & 0.13        & 0.04       & 0.11        & 0.01       & 0.00        & 0.00        & 0.17        & 0.00        & \textbf{0.77} & \textbf{0.72} \\
2002  & 0.14        & 0.04       & 0.11        & 0.01       & 0.00        & 0.00        & 0.29        & 0.01        & \textbf{0.82} & \textbf{0.80} \\
2003  & 0.09        & 0.02       & 0.20        & 0.02       & 0.00        & 0.00        & 0.08        & 0.02        & \textbf{0.83} & \textbf{0.79} \\
2004  & 0.11        & 0.05       & 0.06        & 0.00       & 0.00        & 0.00        & 0.00        & 0.00        & \textbf{0.84} & \textbf{0.68} \\
2005  & 0.07        & 0.11       & 0.06        & 0.01       & 0.00        & 0.00        & 0.00        & 0.00        & \textbf{0.71} & \textbf{0.66} \\
2006  & 0.01        & 0.01       & 0.07        & 0.01       & 0.00        & 0.00        & 0.00        & 0.00        & \textbf{0.43} & \textbf{0.51} \\
2007  & 0.01        & 0.08       & 0.00        & 0.00       & 0.00        & 0.00        & 0.00        & 0.00        & \textbf{0.28} & \textbf{0.44} \\
2008  & 0.01        & 0.04       & 0.00        & 0.00       & 0.00        & 0.00        & 0.00        & 0.00        & \textbf{0.15} & \textbf{0.33} \\
2009  & 0.00        & 0.00       & 0.00        & 0.00       & 0.00        & 0.00        & 0.00        & 0.00        & \textbf{0.09} & \textbf{0.76} \\
\bottomrule
\end{tabular}
\end{table}

We evaluated Augur against other four state-of-the-art algorithms: Fast Greedy (FG), Leading Eigenvector (LE), Fuzzy C-Means (FCM), and Clique Percolation Method (CPM).
The evaluation has been performed across time, and as shown in Table~\ref{tab:augur-per}, Augur outperformed the alternative approaches in terms of both precision and recall in all years. Further details of Augur, the gold standard and the evaluation are available in Salatino et al.~\cite{salatino2018b}.

\subsection{The ResearchFlow Forecaster}
\label{forecast-rf}
In this section, we show how the ResearchFlow approach introduced in Section~\ref{analysis-rf} can also support a very accurate forecaster to predict the impact of research topics in industry.

A good measure to assess the impact of research trends on commercial organisations is the number of relevant patents granted to companies. 
The literature proposes a range of approaches for this task~\cite{altuntas2015analysis,ramadhan2018artificial}. However, most of these methods focus only on patents since they are limited by current datasets that do not typically integrate research articles nor can they distinguish between documents produced by academia or industry.
We thus hypothesised that a scientific knowledge graph like AIDA, which integrates all the information about publications and patents and their origin, should offer a richer set of features, ultimately yielding a better performance in comparison to approaches that rely solely on the number of publications or patents.

In order to train a forecaster, we created a gold standard, in which, for each topic in CSO, we selected all the timeframes of five years in which the topic had not yet emerged (less than ten patents). We then labelled each of these samples as \textit{True} if the topic produced more than 50 industrial patents in the following ten years and \textit{False} otherwise. The resulting dataset includes 9,776 labelled samples, each composed of four time series: 
\begin{inline}
  \item papers from academia \textbf{RA}, 
  \item papers from industry \textbf{RI}, 
  \item patents from academia \textbf{PA},
  \item patents from industry \textbf{PI}.
\end{inline}
We then trained five machine learning classifiers on the gold standard: Logistic Regression (\textbf{LR}), Random Forest (\textbf{RF}), AdaBoost (\textbf{AB}), Convoluted Neural Network (\textbf{CNN}), and Long Short-term Memory Neural Network (\textbf{LSTM}). 
We ran each of them on 17 possible combinations of the four time series in order to assess which set of features would yield the best results. \textbf{RA-RI} concatenates the time series of research papers in academia (\textbf{RA}) and industry (\textbf{RI}) together. Instead, R and P sum the contribution of all papers and patents, respectively. Finally, \textbf{RA-RI-PA-PI} concatenates the four time series.
performed a 10-fold cross-validation of the data and measured the performance of the
classifiers by computing the average precision (\textbf{P}), recall (\textbf{R}), and F1 (\textbf{F}).

\begin{table}[t]
\caption{\label{tab:forecast} Performance of the five classifiers on 17 combinations of time series. In bold the best F1 (F) for each combination. The table and the experiments were previously reported in~\cite{salatino2020b}.}
\begin{tabular}{l|l|l|l|l|l|l|l|l|l|l|l|l|l|l|l|}
\toprule
\multicolumn{1}{c|}{} & \multicolumn{3}{c|}{LR} & \multicolumn{3}{c|}{RF} & \multicolumn{3}{c|}{AB} & \multicolumn{3}{c|}{CNN} & \multicolumn{3}{c|}{LSTM} \\
               & P\%.   & R\%   & F\%   & P\%    & R\%   & F\%   & P\%    & R\%   & F\%   & P\%    & R\%    & F\%   & P\%    & R\%    & F\%    \\
\midrule
RA             & 70.8   & 45.2  & 55.2  & 63.3   & 55.8  & 59.2  & 66.0   & 58.4  & 61.9  & 64.1   & 66.3   & \textbf{65.0}  & 65.2   & 64.2   & 64.6   \\
RI             & 83.5   & 67.1  & 74.4  & 78.9   & 69.8  & 74.0  & 80.0   & 73.1  & 76.4  & 79.2   & 75.1   & \textbf{77.0}  & 79.1   & 74.8   & 76.9   \\
PA             & 58.3   & 15.3  & 24.2  & 60.4   & 15.4  & 24.5  & 59.3   & 16.0  & \textbf{25.2}  & 60.5   & 15.7   & 24.9  & 60.8   & 15.6   & 24.8   \\
PI             & 76.5   & 69.0  & 72.5  & 73.9   & 68.4  & 71.0  & 75.6   & 71.8  & 73.6  & 73.7   & 76.6   & 75.0  & 74.1   & 76.6   & \textbf{75.2}   \\
R              & 73.7   & 48.8  & 58.7  & 65.5   & 59.7  & 62.5  & 68.6   & 63.1  & 65.6  & 67.6   & 69.2   & \textbf{68.3}  & 67.2   & 69.4   & 68.2   \\
P              & 76.5   & 68.6  & 72.3  & 72.8   & 67.6  & 70.0  & 74.4   & 71.6  & 73.0  & 73.2   & 76.1   & 74.6  & 73.1   & 76.6   & \textbf{74.8}   \\
\midrule
RA, RI         & 85.7   & 70.9  & 77.6  & 80.5   & 76.0  & 78.2  & 82.6   & 76.6  & 79.5  & 78.9   & 75.1   & 76.8  & 82.2   & 79.3   & \textbf{80.7}   \\
RA, PA         & 70.3   & 47.0  & 56.3  & 63.1   & 55.5  & 59.0  & 66.5   & 59.3  & 62.6  & 64.5   & 65.1   & 64.5  & 65.4   & 64.2   & \textbf{64.6}   \\
RA, PI         & 79.6   & 73.7  & 76.5  & 77.2   & 74.3  & 75.7  & 79.1   & 76.5  & 77.7  & 75.2   & 76.3   & 75.7  & 77.4   & 81.9   & \textbf{79.5}   \\
RI, PA         & 83.3   & 67.0  & 74.3  & 77.9   & 70.8  & 74.1  & 79.6   & 73.0  & 76.1  & 78.6   & 75.6   & 77.0  & 79.1   & 75.2   & \textbf{77.1}   \\
RI, PI         & 83.4   & 77.3  & 80.2  & 81.0   & 77.3  & 79.1  & 82.7   & 78.6  & 80.6  & 82.0   & 78.6   & 80.2  & 81.7   & 81.2   & \textbf{81.4}   \\
PA, PI         & 76.7   & 68.6  & 72.4  & 74.2   & 69.0  & 71.5  & 75.9   & 71.5  & 73.6  & 71.1   & 70.8   & 70.9  & 73.8   & 76.7   & \textbf{75.2}   \\
\midrule
RA, RI, PA     & 85.2   & 71.4  & 77.7  & 80.8   & 75.4  & 78.0  & 82.5   & 77.0  & 79.6  & 82.6   & 78.1   & \textbf{80.3}  & 82.6   & 78.2   & \textbf{80.3}   \\
RA, RI, PI     & 85.4   & 79.8  & 82.5  & 84.5   & 80.5  & 82.4  & 84.6   & 81.2  & 82.9  & 83.8   & 84.7   & 84.2  & 84.1   & 85.4   & \textbf{84.7}   \\
RA, PA, PI     & 79.6   & 73.9  & 76.6  & 77.5   & 74.4  & 75.9  & 79.2   & 76.5  & 77.8  & 78.9   & 78.6   & 78.6  & 77.4   & 81.4   & \textbf{79.2}   \\
RI, PA, PI     & 83.6   & 77.5  & 80.4  & 81.1   & 78.0  & 79.5  & 82.7   & 78.6  & 80.6  & 82.2   & 80.9   & \textbf{81.5}  & 81.1   & 81.0   & 81.1   \\
\midrule
RA, RI, PA, PI & 85.4   & 79.8  & 82.5  & 83.8   & 80.0  & 81.8  & 84.6   & 81.2  & 82.9  & 84.7   & 81.3   & 82.9  & 83.2   & 86.1   & \textbf{84.6}   \\
\bottomrule
\end{tabular}
\end{table}

Table~\ref{tab:forecast} reports the results of our experiment. LSTM outperforms all the other solutions, yielding the highest F1 for 12 of the 17 feature combinations and the highest average F1 (73.7\%). CNN (72.8\%) and AB (72.3\%) also produce competitive results.  

As hypothesised, taking advantage of the full set of features available in AIDA – broken into its times series – significantly outperforms (with F1 = 84.6\%) the version which uses only the number of patents by companies (75.2\%). 
Considering the origin (academia and industry) of the publications and the patents also increases performance: RA-RI (80.7\%) significantly outperforms R (68.2\%) and PA-PI (75.2\%) is marginally better than P (74.8\%). 

In conclusion, the experiments confirm that the granular representation of research topics in AIDA can effectively support deep learning approaches for forecasting the impact of research topics on the industrial sector. It also validates the intuition that including features from research articles can be very useful when predicting industrial trends.

\subsection{Predicting the spreading of technologies}
\label{forecast-techonologies}


Every new piece of research, no matter how ground-breaking,
adopts previous knowledge and reuses tools and methodologies.
Typically, a technology is expected to originate in the context of a research area and then spread and contribute to several other fields.
Unfortunately, given the tremendous amount of knowledge produced yearly, it is difficult to digest it and pick up on an interesting piece of research from a different field. 
Thus, the transfer of a technology from one research area (e.g., Semantic Web) to a different, and possibly conceptually distant, one (e.g., Digital Humanities) may take several years, slowing down the whole research progress.
It is, therefore, crucial to be able to anticipate technology spreading across the different research community. 

The Technology-Topic Framework (TTF)~\cite{osborne2017ttf} is a recent approach that uses a semantically enhanced technology-topic model to predict the technologies that will be adopted by a research field. It aims at suggesting promising technologies to scholars from a field, thus helping to accelerate the knowledge flow and the pace of technology propagation.
It is based on the hypothesis that \textit{technologies that exhibit similar spreading patterns across multiple research topics will tend to be adopted by similar topics}. It thus characterises technologies in terms of a set of topics drawn from a large-scale ontology of research areas over a given time period and applies machine learning on these data so to forecast technology spreading. 

The prototype of TTF takes as input three knowledge bases:
\begin{inline}
    \item a dataset of 16 million research papers from 1990 to 2013 mainly in the field of Computer Science, described by means of their titles, abstracts, and keywords;
    \item a list of 1,118 input technologies extracted automatically from paper abstracts and from generic KBs (e.g., DBpedia~\cite{auer2007dbpedia}) and then manually curated, which are associated to at least 10 relevant publications in the research paper dataset;
    \item 173 relevant research topics and their relationships drawn from the Computer Science Ontology~\cite{salatino2020,salatino2018}.
\end{inline}
TTF generates a 3-dimensional co-occurrence matrix that tracks the number of papers in which a given technology is associated with a given topic in a given year. 
It then uses this matrix to train a classifier on a technology recent history in order to predict its future course. 
The Random Forest classifier performed best on this data, obtaining a precision of 74.4\% and a recall of 47.7\%, outperforming alternative solutions such as Logistic Regression, Decision Tree, Gradient Boosting, SVM, and a simple Neural Network (see~\cite{osborne2017ttf} for the full details).


\begin{table}[]
\caption{\label{tab:forecast-tech} Performance of Random Forest on the first 24 topics, with at least 50 positive labels, ordered by F1 score. The table and the experiments were previously reported in~\cite{osborne2017ttf}.}
\begin{tabular}{l|l|l|l||l|l|l|l|}
\hline
\textbf{Topics}         & \textbf{Prec. \%} & \textbf{Rec. \%} & \textbf{F1 \%} & \textbf{Topics}        & \textbf{Prec. \%} & \textbf{Rec. \%} & \textbf{F1 \%} \\ \hline
information retrieval   & 92.6        & 66.8       & 77.6     & wireless networks      & 64.7        & 47.8       & 55.0     \\ \hline
database systems        & 82.6        & 65.9       & 73.3     & sensor networks        & 71.9        & 43.6       & 54.3     \\ \hline
world wide web          & 88.6        & 56.1       & 68.7     & software engineering   & 70.6        & 44.0       & 54.2     \\ \hline
artificial intelligence & 83.6        & 55.2       & 66.5     & distributed comp. sys. & 67.5        & 45.0       & 54.0     \\ \hline
computer architecture   & 68.3        & 63.3       & 65.7     & quality of service     & 59.6        & 48.6       & 53.5     \\ \hline
computer networks       & 82.1        & 54.0       & 65.2     & imaging systems        & 100.0       & 35.8       & 52.8     \\ \hline
image coding            & 96.8        & 46.9       & 63.2     & data mining            & 60.8        & 45.3       & 52.0     \\ \hline
P2P networks            & 78.9        & 50.8       & 61.9     & computer vision        & 92.3        & 36.0       & 51.8     \\ \hline
Telecom. traffic        & 70.8        & 48.1       & 57.3     & programming languages  & 65.3        & 42.0       & 51.2     \\ \hline
wireless telecom. sys.  & 74.4        & 46.4       & 57.1     & problem solving        & 69.0        & 39.7       & 50.4     \\ \hline
sensors                 & 78.8        & 43.7       & 56.2     & semantic web           & 77.8        & 37.1       & 50.2     \\ \hline
web services            & 83.3        & 42.2       & 56.0     & image quality          & 74.2        & 37.7       & 50.0     \\ \hline
\end{tabular}
\end{table}

Table~\ref{tab:forecast-tech} reports the results of the version of TTF using Random Forest on the most frequent 24 topics in the experiment. 
We can see that TTF obtains the best performance on topics associated with a large set of publications and technologies, such as Information Retrieval and Databases. 
However, it is also able to obtain satisfactory results in many other fields, especially the ones that are usually associated with a coherent set of technologies, such as Computer Networks, Sensors, and Peer-to-peer Systems. 
This confirms the hypothesis that is indeed possible to learn from historical spreading patterns and forecast technology propagation, at least for the set of topics that are more involved in technology propagation events.

\section{Conclusion}
\label{conclusions}
In this chapter, we presented a framework for detecting, analysing, and predicting research topics based on a large-scale knowledge graph characterising research articles according to the research topics from the Computer Science Ontology (CSO)~\cite{salatino2020,salatino2018}.
We first illustrated how to 
annotate a scientific knowledge graph describing research articles and their metadata with a set of research topics from a domain ontology. 
We then discussed several methods that build on this knowledge graph for analysing research from different perspectives. 
Finally, we presented two approaches for predicting research trends based on this framework. 

%
%
\bibliographystyle{splncs04}
\bibliography{bibliography}

\begin{thebibliography}{10}
\providecommand{\url}[1]{\texttt{#1}}
\providecommand{\urlprefix}{URL }
\providecommand{\doi}[1]{https://doi.org/#1}

\bibitem{altuntas2015analysis}
Altuntas, S., Dereli, T., Kusiak, A.: Analysis of patent documents with
  weighted association rules. Technological Forecasting and Social Change
  \textbf{92},  249--262 (2015). \doi{10.1016/j.techfore.2014.09.012}

\bibitem{angioni2020}
Angioni, S., Salatino, A., Osborne, F., Recupero, D.R., Motta, E.: The aida
  dashboard: Analysing conferences with semantic technologies. In: 19th
  International Semantic Web Conference (ISWC 2020) (2020),
  \url{http://oro.open.ac.uk/72293/}

\bibitem{angioni2020a}
Angioni, S., Salatino, A.A., Osborne, F., Recupero, D.R., Motta, E.:
  Integrating knowledge graphs for analysing academia and industry dynamics.
  In: Bellatreche, L., Bielikov{\'a}, M., Boussa{\"i}d, O., Catania, B.,
  Darmont, J., Demidova, E., Duchateau, F., Hall, M., Mer{\v{c}}un, T.,
  Novikov, B., Papatheodorou, C., Risse, T., Romero, O., Sautot, L., Talens,
  G., Wrembel, R., {\v{Z}}umer, M. (eds.) ADBIS, TPDL and EDA 2020 Common
  Workshops and Doctoral Consortium. pp. 219--225. Springer International
  Publishing, Cham (2020)

\bibitem{aromataris2014constructing}
Aromataris, E., Riitano, D.: Constructing a search strategy and searching for
  evidence. American Journal of Nursing  \textbf{114}(5),  49--56 (2014).
  \doi{10.1097/01.NAJ.0000446779.99522.f6}

\bibitem{auer2007dbpedia}
Auer, S., Bizer, C., Kobilarov, G., Lehmann, J., Cyganiak, R., Ives, Z.:
  Dbpedia: A nucleus for a web of open data. In: The semantic web, pp.
  722--735. Springer (2007). \doi{10.1007/978-3-540-76298-0\_52}

\bibitem{augenstein2012lodifier}
Augenstein, I., Pad{\'o}, S., Rudolph, S.: Lodifier: Generating linked data
  from unstructured text. In: Extended Semantic Web Conference. pp. 210--224.
  Springer (2012). \doi{10.1007/978-3-642-30284-8\_21}

\bibitem{beck2020automatic}
Beck, M., Rizvi, S.T.R., Dengel, A., Ahmed, S.: From automatic keyword
  detection to ontology-based topic modeling. In: International Workshop on
  Document Analysis Systems. pp. 451--465. Springer (2020).
  \doi{10.1007/978-3-030-57058-3\_32}

\bibitem{blei2003}
Blei, D.M., Ng, A.Y., Jordan, M.I.: Latent dirichlet allocation. J. Mach.
  Learn. Res.  \textbf{3}(null),  993–1022 (Mar 2003)

\bibitem{bolelli2009}
Bolelli, L., Ertekin, {\c{S}}., Giles, C.L.: Topic and trend detection in text
  collections using latent dirichlet allocation. In: Boughanem, M., Berrut, C.,
  Mothe, J., Soule-Dupuy, C. (eds.) Advances in Information Retrieval. pp.
  776--780. Springer Berlin Heidelberg, Berlin, Heidelberg (2009).
  \doi{10.1007/978-3-642-00958-7\_84}

\bibitem{borges2019semantic}
Borges, M.V.M., dos Reis, J.C.: Semantic-enhanced recommendation of video
  lectures. In: 2019 IEEE 19th International Conference on Advanced Learning
  Technologies (ICALT). vol.~2161, pp. 42--46. IEEE (2019).
  \doi{10.1109/ICALT.2019.00013}

\bibitem{bornmann2015}
Bornmann, L., Mutz, R.: Growth rates of modern science: A bibliometric analysis
  based on the number of publications and cited references. Journal of the
  Association for Information Science and Technology  \textbf{66}(11),
  2215--2222 (2015). \doi{10.1002/asi.23329}

\bibitem{boyack2014}
Boyack, K.W., Klavans, R.: Creation of a highly detailed, dynamic, global model
  and map of science. Journal of the Association for Information Science and
  Technology  \textbf{65}(4),  670--685 (2014). \doi{10.1002/asi.22990}

\bibitem{boyack2005}
Boyack, K.W., Klavans, R., B{\"{o}}rner, K.: {Mapping the backbone of science}.
  Scientometrics  \textbf{64}(3),  351--374 (2005).
  \doi{10.1007/s11192-005-0255-6}

\bibitem{cano2016}
Cano-Basave, A.E., Osborne, F., Salatino, A.A.: Ontology forecasting in
  scientific literature: Semantic concepts prediction based on
  innovation-adoption priors. In: Blomqvist, E., Ciancarini, P., Poggi, F.,
  Vitali, F. (eds.) Knowledge Engineering and Knowledge Management. pp. 51--67.
  Springer International Publishing, Cham (2016)

\bibitem{cena2013anisotropic}
Cena, F., Likavec, S., Osborne, F.: Anisotropic propagation of user interests
  in ontology-based user models. Information Sciences  \textbf{250},  40--60
  (2013). \doi{10.1016/j.ins.2013.07.006}

\bibitem{chatzopoulos2020artsim}
Chatzopoulos, S., Vergoulis, T., Kanellos, I., Dalamagas, T., Tryfonopoulos,
  C.: Artsim: improved estimation of current impact for recent articles. In:
  ADBIS, TPDL and EDA 2020 Common Workshops and Doctoral Consortium. pp.
  323--334. Springer (2020). \doi{10.1007/978-3-030-55814-7\_27}

\bibitem{chicaiza2020using}
Chicaiza, J., Re{\'a}tegui, R.: Using domain ontologies for text
  classification. a use case to classify computer science papers. In:
  Iberoamerican Knowledge Graphs and Semantic Web Conference. pp. 166--180.
  Springer (2020). \doi{10.1007/978-3-030-65384-2\_13}

\bibitem{cimiano2005text2onto}
Cimiano, P., V{\"o}lker, J.: Text2onto. In: International Conference on
  Application of Natural Language to Information Systems. pp. 227--238.
  Springer (2005). \doi{10.1007/11428817\_21}

\bibitem{decker2007}
Decker, S.L.: Detection of bursty and emerging trends towards identification of
  researchers at the early stage of trends (2007)

\bibitem{dessi2021generating}
Dess{\`\i}, D., Osborne, F., Recupero, D.R., Buscaldi, D., Motta, E.:
  Generating knowledge graphs by employing natural language processing and
  machine learning techniques within the scholarly domain. Future Generation
  Computer Systems  \textbf{116},  253--264 (2021)

\bibitem{dessi2020ai}
Dess{\`\i}, D., Osborne, F., Recupero, D.R., Buscaldi, D., Motta, E., Sack, H.:
  Ai-kg: an automatically generated knowledge graph of artificial intelligence.
  In: International Semantic Web Conference. pp. 127--143. Springer (2020).
  \doi{10.1007/978-3-030-62466-8\_9}

\bibitem{dicaro2017}
{Di Caro}, L., Guerzoni, M., Nuccio, M., Siragusa, G.: {A Bimodal Network
  Approach to Model Topic Dynamics}  (2017),
  \url{https://arxiv.org/pdf/1709.09373.pdf}

\bibitem{duvvuru2013}
Duvvuru, A., Radhakrishnan, S., More, D., Kamarthi, S., Sultornsanee, S.:
  Analyzing structural \& temporal characteristics of keyword system in
  academic research articles. Procedia Computer Science  \textbf{20},  439--445
  (2013). \doi{10.1016/j.procs.2013.09.300}, complex Adaptive Systems

\bibitem{erten2004}
Erten, C., Harding, P.J., Kobourov, S.G., Wampler, K., Yee, G.: {Exploring the
  computing literature using temporal graph visualization}. Visualization and
  Data Analysis 2004  \textbf{5295},  45--56 (2004). \doi{10.1117/12.539245}

\bibitem{fenner2019introducing}
Fenner, M., Aryani, A.: Introducing the pid graph (2019).
  \doi{10.5438/jwvf-8a66}

\bibitem{gangemi2017semantic}
Gangemi, A., Presutti, V., Reforgiato~Recupero, D., Nuzzolese, A.G., Draicchio,
  F., Mongiov{\`\i}, M.: Semantic web machine reading with fred. Semantic Web
  \textbf{8}(6),  873--893 (2017). \doi{10.3233/SW-160240}

\bibitem{garciasilva2021}
Garcia-Silva, A., Gomez-Perez, J.M.: Classifying scientific publications with
  bert -- is self-attention a feature selection method? (2021)

\bibitem{griffiths2004}
Griffiths, T.L., Steyvers, M.: Finding scientific topics. Proceedings of the
  National Academy of Sciences  \textbf{101}(suppl 1),  5228--5235 (2004).
  \doi{10.1073/pnas.0307752101}

\bibitem{he2009}
He, Q., Chen, B., Giles, C.L.: {Detecting Topic Evolution in Scientific
  Literature : How Can Citations Help ?} Cikm pp. 957--966 (2009).
  \doi{10.1145/1645953.1646076}

\bibitem{hogan2020knowledge}
Hogan, A., Blomqvist, E., Cochez, M., d'Amato, C., de~Melo, G., Gutierrez, C.,
  Gayo, J.E.L., Kirrane, S., Neumaier, S., Polleres, A., et~al.: Knowledge
  graphs. arXiv preprint arXiv:2003.02320  (2020)

\bibitem{jaradeh2019open}
Jaradeh, M.Y., Oelen, A., Farfar, K.E., Prinz, M., D'Souza, J., Kismih\'{o}k,
  G., Stocker, M., Auer, S.: Open research knowledge graph: Next generation
  infrastructure for semantic scholarly knowledge. In: Proceedings of the 10th
  International Conference on Knowledge Capture. p. 243–246. K-CAP '19,
  Association for Computing Machinery, New York, NY, USA (2019).
  \doi{10.1145/3360901.3364435}

\bibitem{jo2007}
Jo, Y., Lagoze, C., Giles, C.L.: Detecting research topics via the correlation
  between graphs and texts. In: Proceedings of the 13th ACM SIGKDD
  International Conference on Knowledge Discovery and Data Mining. p.
  370–379. KDD '07, Association for Computing Machinery, New York, NY, USA
  (2007). \doi{10.1145/1281192.1281234}

\bibitem{kandimalla2020}
Kandimalla, B., Rohatgi, S., Wu, J., Giles, C.L.: Large scale subject category
  classification of scholarly papers with deep attentive neural networks.
  Frontiers in Research Metrics and Analytics  \textbf{5}, ~31 (2021).
  \doi{10.3389/frma.2020.600382}

\bibitem{kirrane2020decade}
Kirrane, S., Sabou, M., Fern{\'a}ndez, J.D., Osborne, F., Robin, C., Buitelaar,
  P., Motta, E., Polleres, A.: A decade of semantic web research through the
  lenses of a mixed methods approach. Semantic Web  \textbf{11}(6),  979--1005
  (2020). \doi{10.3233/SW-200371}

\bibitem{kitchenhamTR2004}
Kitchenham, B.: Procedures for performing systematic reviews. Keele, UK, Keele
  University  \textbf{33}(2004),  1--26 (2004)

\bibitem{kitchenham2007guidelines}
Kitchenham, B.A., Charters, S.: Guidelines for performing systematic literature
  reviews in software engineering (2007)

\bibitem{krampen2011}
Krampen, G., von Eye, A., Schui, G.: {Forecasting trends of development of
  psychology from a bibliometric perspective}. Scientometrics  \textbf{87},
  687--694 (2011). \doi{10.1007/s11192-011-0357-2}

\bibitem{LaBruzzo2018}
{La Bruzzo}, S., Manghi, P., Mannocci, A.: {OpenAIRE's DOIBoost - Boosting
  Crossref for Research}. In: Manghi, P., Candela, L., Silvello, G. (eds.)
  Digital Libraries: Supporting Open Science. pp. 133--143. Springer
  International Publishing, Cham (jan 2019).
  \doi{10.1007/978-3-030-11226-4\_11}

\bibitem{leydesdorff2013}
Leydesdorff, L., Rafols, I., Chen, C.: Interactive overlays of journals and the
  measurement of interdisciplinarity on the basis of aggregated
  journal–journal citations. Journal of the American Society for Information
  Science and Technology  \textbf{64}(12),  2573--2586 (2013).
  \doi{10.1002/asi.22946}

\bibitem{likavec2015property}
Likavec, S., Osborne, F., Cena, F.: Property-based semantic similarity and
  relatedness for improving recommendation accuracy and diversity.
  International Journal on Semantic Web and Information Systems (IJSWIS)
  \textbf{11}(4),  1--40 (2015). \doi{10.4018/IJSWIS.2015100101}

\bibitem{liu2012automatic}
Liu, X., Song, Y., Liu, S., Wang, H.: Automatic taxonomy construction from
  keywords. In: Proceedings of the 18th ACM SIGKDD international conference on
  Knowledge discovery and data mining. pp. 1433--1441. ACM (2012).
  \doi{10.1145/2339530.2339754}

\bibitem{ScholarLensViz2020}
L\"offler, F., Wesp, V., Babalou, S., Kahn, P., Lachmann, R., Sateli, B.,
  Witte, R., K\"onig-Ries, B.: Scholarlensviz: A visualization framework for
  transparency in semantic user profiles. In: Taylor, K., Gonçalves, R.,
  Lecue, F., Yan, J. (eds.) Proceedings of the ISWC 2020 Demos and Industry
  Tracks: From Novel Ideas to Industrial Practice co-located with 19th
  International Semantic Web Conference (ISWC 2020), Globally online, November
  1-6, 2020 (UTC). (2020)

\bibitem{lulaadvanced}
Lula, P., Dospinescu, O., Homocianu, D., Sireteanu, N.A.: An advanced analysis
  of cloud computing concepts based on the computer science ontology.
  Computers, Materials \& Continua  \textbf{66}(3),  2425--2443 (2021).
  \doi{10.32604/cmc.2021.013771}

\bibitem{lv2011}
Lv, P.H., Wang, G.F., Wan, Y., Liu, J., Liu, Q., Ma, F.C.: {Bibliometric trend
  analysis on global graphene research}. Scientometrics  \textbf{88},  399--419
  (2011). \doi{10.1007/s11192-011-0386-x}

\bibitem{mai2018}
Mai, F., Galke, L., Scherp, A.: Using deep learning for title-based semantic
  subject indexing to reach competitive performance to full-text. In:
  Proceedings of the 18th ACM/IEEE on Joint Conference on Digital Libraries. p.
  169–178. JCDL '18, Association for Computing Machinery, New York, NY, USA
  (2018). \doi{10.1145/3197026.3197039}

\bibitem{Manghi2020}
Manghi, P., Atzori, C., Bardi, A., Baglioni, M., Schirrwagen, J.,
  Dimitropoulos, H., La~Bruzzo, S., Foufoulas, I., Löhden, A., Bäcker, A.,
  Mannocci, A., Horst, M., Jacewicz, P., Czerniak, A., Kiatropoulou, K.,
  Kokogiannaki, A., De~Bonis, M., Artini, M., Ottonello, E., Lempesis, A.,
  Ioannidis, A., Manola, N., Principe, P.: Openaire research graph dump (Nov
  2020). \doi{10.5281/zenodo.4279381}

\bibitem{mannocci2019evolution}
Mannocci, A., Osborne, F., Motta, E.: The evolution of ijhcs and chi: A
  quantitative analysis. International Journal of Human-Computer Studies
  \textbf{131},  23--40 (2019). \doi{10.1016/j.ijhcs.2019.05.009}

\bibitem{mikolov2013}
Mikolov, T., Chen, K., Corrado, G., Dean, J.: Efficient estimation of word
  representations in vector space (2013)

\bibitem{nayyeri2021trans4e}
Nayyeri, M., Cil, G.M., Vahdati, S., Osborne, F., Rahman, M., Angioni, S.,
  Salatino, A., Recupero, D.R., Vassilyeva, N., Motta, E., Lehmann, J.:
  Trans4e: Link prediction on scholarly knowledge graphs. Neurocomputing
  (2021)

\bibitem{Nuzzolese2016}
Nuzzolese, A.G., Gentile, A.L., Presutti, V., Gangemi, A.: {Conference linked
  data: The scholarlydata project}. In: Lecture Notes in Computer Science
  (including subseries Lecture Notes in Artificial Intelligence and Lecture
  Notes in Bioinformatics). vol. 9982 LNCS, pp. 150--158 (2016).
  \doi{10.1007/978-3-319-46547-0\_16}

\bibitem{osborne2016techminer}
Osborne, F., De~Ribaupierre, H., Motta, E.: Techminer: extracting technologies
  from academic publications. In: European Knowledge Acquisition Workshop. pp.
  463--479. Springer (2016)

\bibitem{osborne2017ttf}
Osborne, F., Mannocci, A., Motta, E.: Forecasting the spreading of technologies
  in research communities. In: Proceedings of the Knowledge Capture Conference.
  K-CAP 2017, Association for Computing Machinery, New York, NY, USA (2017).
  \doi{10.1145/3148011.3148030}

\bibitem{osborne2015}
Osborne, F., Motta, E.: Klink-2: Integrating multiple web sources to generate
  semantic topic networks. In: Arenas, M., Corcho, O., Simperl, E., Strohmaier,
  M., d'Aquin, M., Srinivas, K., Groth, P., Dumontier, M., Heflin, J.,
  Thirunarayan, K., Thirunarayan, K., Staab, S. (eds.) The Semantic Web - ISWC
  2015. pp. 408--424. Springer International Publishing, Cham (2015).
  \doi{10.1007/978-3-319-25007-6\_24}

\bibitem{osborne2018pragmatic}
Osborne, F., Motta, E.: Pragmatic ontology evolution: reconciling user
  requirements and application performance. In: International Semantic Web
  Conference. pp. 495--512. Springer (2018)

\bibitem{osborne2013}
Osborne, F., Motta, E., Mulholland, P.: Exploring scholarly data with rexplore.
  In: Alani, H., Kagal, L., Fokoue, A., Groth, P., Biemann, C., Parreira, J.X.,
  Aroyo, L., Noy, N., Welty, C., Janowicz, K. (eds.) The Semantic Web -- ISWC
  2013. pp. 460--477. Springer Berlin Heidelberg, Berlin, Heidelberg (2013).
  \doi{10.1007/978-3-642-41335-3\_29}

\bibitem{osborne2019reducing}
Osborne, F., Muccini, H., Lago, P., Motta, E.: Reducing the effort for
  systematic reviews in software engineering. Data Science  \textbf{2}(1-2),
  311--340 (2019). \doi{10.3233/DS-190019}

\bibitem{osborne2016}
Osborne, F., Salatino, A., Birukou, A., Motta, E.: Automatic classification of
  springer nature proceedings with smart topic miner. In: Groth, P., Simperl,
  E., Gray, A., Sabou, M., Kr{\"o}tzsch, M., Lecue, F., Fl{\"o}ck, F., Gil, Y.
  (eds.) The Semantic Web -- ISWC 2016. pp. 383--399. Springer International
  Publishing, Cham (2016)

\bibitem{osborne2014hybrid}
Osborne, F., Scavo, G., Motta, E.: A hybrid semantic approach to building
  dynamic maps of research communities. In: International Conference on
  Knowledge Engineering and Knowledge Management. pp. 356--372. Springer (2014)

\bibitem{osborne2014}
Osborne, F., Scavo, G., Motta, E.: Identifying diachronic topic-based research
  communities by clustering shared research trajectories. In: Presutti, V.,
  d'Amato, C., Gandon, F., d'Aquin, M., Staab, S., Tordai, A. (eds.) The
  Semantic Web: Trends and Challenges. pp. 114--129. Springer International
  Publishing, Cham (2014)

\bibitem{peroni2017research}
Peroni, S., Osborne, F., Di~Iorio, A., Nuzzolese, A.G., Poggi, F., Vitali, F.,
  Motta, E.: Research articles in simplified html: a web-first format for
  html-based scholarly articles. PeerJ Computer Science  \textbf{3}, ~e132
  (2017). \doi{10.7717/peerj-cs.132}

\bibitem{Peroni2020}
Peroni, S., Shotton, D.: {OpenCitations, an infrastructure organization for
  open scholarship}. Quantitative Science Studies  \textbf{1}(1),  428--444
  (2020). \doi{10.1162/qss{\_}a{\_}00023}

\bibitem{petersen2015guidelines}
Petersen, K., Vakkalanka, S., Kuzniarz, L.: Guidelines for conducting
  systematic mapping studies in software engineering: An update. Information
  and Software Technology  \textbf{64},  1--18 (2015).
  \doi{10.1016/j.infsof.2015.03.007}

\bibitem{petrucci2016ontology}
Petrucci, G., Ghidini, C., Rospocher, M.: Ontology learning in the deep. In:
  European Knowledge Acquisition Workshop. pp. 480--495. Springer (2016).
  \doi{10.1007/978-3-319-49004-5\_31}

\bibitem{ramadhan2018artificial}
Ramadhan, M.H., Malik, V.I., Sjafrizal, T.: Artificial neural network approach
  for technology life cycle construction on patent data. In: 2018 5th
  International Conference on Industrial Engineering and Applications (ICIEA).
  pp. 499--503 (2018). \doi{10.1109/IEA.2018.8387152}

\bibitem{rossanez2020representing}
Rossanez, A., dos Reis, J.C., da~Silva~Torres, R.: Representing scientific
  literature evolution via temporal knowledge graphs  (2020)

\bibitem{salatino2020b}
Salatino, A., Osborne, F., Motta, E.: Researchflow: Understanding the knowledge
  flow between academia and industry. In: Keet, C.M., Dumontier, M. (eds.)
  Knowledge Engineering and Knowledge Management. pp. 219--236. Springer
  International Publishing, Cham (2020). \doi{10.1007/978-3-030-61244-3\_16}

\bibitem{salatino2018classifying}
Salatino, A., Thanapalasingam, T., Mannocci, A., Osborne, F., Motta, E.:
  Classifying research papers with the computer science ontology  (2018)

\bibitem{salatino2019a}
Salatino, A.A., Osborne, F., Birukou, A., Motta, E.: Improving editorial
  workflow and metadata quality at springer nature. In: Ghidini, C., Hartig,
  O., Maleshkova, M., Sv{\'a}tek, V., Cruz, I., Hogan, A., Song, J.,
  Lefran{\c{c}}ois, M., Gandon, F. (eds.) The Semantic Web -- ISWC 2019. pp.
  507--525. Springer International Publishing, Cham (2019)

\bibitem{salatino2017}
Salatino, A.A., Osborne, F., Motta, E.: How are topics born? understanding the
  research dynamics preceding the emergence of new areas. PeerJ Computer
  Science  \textbf{3}, ~e119 (2017). \doi{10.7717/peerj-cs.119}

\bibitem{salatino2018b}
Salatino, A.A., Osborne, F., Motta, E.: Augur: Forecasting the emergence of new
  research topics. In: Proceedings of the 18th ACM/IEEE on Joint Conference on
  Digital Libraries. p. 303–312. JCDL '18, Association for Computing
  Machinery, New York, NY, USA (2018). \doi{10.1145/3197026.3197052}

\bibitem{salatino2019}
Salatino, A.A., Osborne, F., Thanapalasingam, T., Motta, E.: The cso
  classifier: Ontology-driven detection of research topics in scholarly
  articles. In: Doucet, A., Isaac, A., Golub, K., Aalberg, T., Jatowt, A.
  (eds.) Digital Libraries for Open Knowledge. pp. 296--311. Springer
  International Publishing, Cham (2019). \doi{10.1007/978-3-030-30760-8\_26}

\bibitem{salatino2020}
Salatino, A.A., Thanapalasingam, T., Mannocci, A., Birukou, A., Osborne, F.,
  Motta, E.: The computer science ontology: A comprehensive
  automatically-generated taxonomy of research areas. Data Intelligence
  \textbf{2}(3),  379--416 (2020). \doi{10.1162/dint\_a\_00055}

\bibitem{salatino2018}
Salatino, A.A., Thanapalasingam, T., Mannocci, A., Osborne, F., Motta, E.: The
  computer science ontology: A large-scale taxonomy of research areas. In:
  Vrande{\v{c}}i{\'{c}}, D., Bontcheva, K., Su{\'a}rez-Figueroa, M.C.,
  Presutti, V., Celino, I., Sabou, M., Kaffee, L.A., Simperl, E. (eds.) The
  Semantic Web -- ISWC 2018. pp. 187--205. Springer International Publishing,
  Cham (2018). \doi{10.1007/978-3-030-00668-6\_12}

\bibitem{satopaa2011}
{Satopaa}, V., {Albrecht}, J., {Irwin}, D., {Raghavan}, B.: Finding a "kneedle"
  in a haystack: Detecting knee points in system behavior. In: 2011 31st
  International Conference on Distributed Computing Systems Workshops. pp.
  166--171 (2011). \doi{10.1109/ICDCSW.2011.20}

\bibitem{sinha2015}
Sinha, A., Shen, Z., Song, Y., Ma, H., Eide, D., Hsu, B.J.P., Wang, K.: An
  overview of microsoft academic service (mas) and applications. In:
  Proceedings of the 24th International Conference on World Wide Web. p.
  243–246. WWW '15 Companion, Association for Computing Machinery, New York,
  NY, USA (2015). \doi{10.1145/2740908.2742839}

\bibitem{Tang2008}
Tang, J., Zhang, J., Yao, L., Li, J., Zhang, L., Su, Z.: {ArnetMiner:
  Extraction and mining of academic social networks}. In: Proceedings of the
  ACM SIGKDD International Conference on Knowledge Discovery and Data Mining.
  pp. 990--998 (2008). \doi{10.1145/1401890.1402008}

\bibitem{thanapalasingam2018}
Thanapalasingam, T., Osborne, F., Birukou, A., Motta, E.: Ontology-based
  recommendation of editorial products. In: Vrande{\v{c}}i{\'{c}}, D.,
  Bontcheva, K., Su{\'a}rez-Figueroa, M.C., Presutti, V., Celino, I., Sabou,
  M., Kaffee, L.A., Simperl, E. (eds.) The Semantic Web -- ISWC 2018. pp.
  341--358. Springer International Publishing, Cham (2018)

\bibitem{tseng2009}
Tseng, Y.H., Lin, Y.I., Lee, Y.Y., Hung, W.C., Lee, C.H.: {A comparison of
  methods for detecting hot topics}. Scientometrics  \textbf{81}(1),  73--90
  (2009). \doi{10.1007/s11192-009-1885-x}

\bibitem{VanRaan2004}
{Van Raan}, A.F.: {Sleeping Beauties in science}. Scientometrics
  \textbf{59}(3),  467--472 (2004). \doi{10.1023/B:SCIE.0000018543.82441.f1}

\bibitem{vergoulis2020veto}
Vergoulis, T., Chatzopoulos, S., Dalamagas, T., Tryfonopoulos, C.: Veto: Expert
  set expansion in academia. In: Hall, M., Mer{\v{c}}un, T., Risse, T.,
  Duchateau, F. (eds.) Digital Libraries for Open Knowledge. pp. 48--61.
  Springer International Publishing, Cham (2020).
  \doi{10.1007/978-3-030-54956-5\_4}

\bibitem{visser2021}
Visser, M., van Eck, N.J., Waltman, L.: {Large-scale comparison of
  bibliographic data sources: Scopus, Web of Science, Dimensions, Crossref, and
  Microsoft Academic} (04 2021). \doi{10.1162/qss\_a\_00112}

\bibitem{Wang2020}
Wang, K., Shen, Z., Huang, C., Wu, C.H., Dong, Y., Kanakia, A.: {Microsoft
  Academic Graph: When experts are not enough}. Quantitative Science Studies
  \textbf{1}(1),  396--413 (2020). \doi{10.1162/qss\_a\_00021}

\bibitem{watatani2013}
Watatani, K., Xie, Z., Nakatsuji, N., Sengoku, S.: {Global competencies of
  regional stem cell research: bibliometrics for investigating and forecasting
  research trends.} Regenerative medicine  \textbf{8}(5),  659--68 (2013).
  \doi{10.2217/rme.13.51}

\bibitem{widodo2011}
{Widodo}, A., {Fanany}, M.I., {Budi}, I.: Technology forecasting in the field
  of apnea from online publications: Time series analysis on latent semantic.
  In: 2011 Sixth International Conference on Digital Information Management.
  pp. 127--132 (2011). \doi{10.1109/ICDIM.2011.6093365}

\bibitem{wohlin2013systematic}
Wohlin, C., Prikladnicki, R.: Systematic literature reviews in software
  engineering. Information and Software Technology  \textbf{55}(6),  919--920
  (2013). \doi{10.1016/j.infsof.2013.02.002}

\bibitem{zhang2021conceptscope}
Zhang, X., Chandrasegaran, S., Ma, K.L.: Conceptscope: Organizing and
  visualizing knowledge in documents based on domain ontology. In: Proceedings
  of the 2021 CHI Conference on Human Factors in Computing Systems. pp. 1--13
  (2021)

\bibitem{zhang2018}
Zhang, Y., Lu, J., Liu, F., Liu, Q., Porter, A., Chen, H., Zhang, G.: Does deep
  learning help topic extraction? a kernel k-means clustering method with word
  embedding. Journal of Informetrics  \textbf{12}(4),  1099--1117 (2018).
  \doi{10.1016/j.joi.2018.09.004}

\end{thebibliography}

\end{document}